\documentclass{article}

\usepackage{arxiv}
\usepackage[utf8]{inputenc} 
\usepackage{url}            
\usepackage{booktabs}       
\usepackage{amsfonts}       
\usepackage{nicefrac}       
\usepackage{microtype}      
\usepackage{lipsum}
\usepackage{amsmath,amsfonts,amssymb,amscd,amsthm,xspace,bm,bbm}
\theoremstyle{plain}

\newtheorem{theorem}{Theorem}

\theoremstyle{definition}

\theoremstyle{remark}

\usepackage[centerlast,small,sc]{caption}
\usepackage{subfigure}
\usepackage{multirow}
\usepackage{txfonts}
\usepackage{enumerate}
\usepackage{graphicx}
\usepackage{algorithm}
\usepackage[noend]{algpseudocode}
\DeclareUnicodeCharacter{FB02}{fl}
\DeclareUnicodeCharacter{FB01}{fi}
\makeatletter
\def\algbackskip{\hskip-\ALG@thistlm}
\makeatother
\DeclareMathOperator*{\argmin}{arg\,min}

\usepackage{cite}

\usepackage{enumitem}
\usepackage{xcolor}


\graphicspath{ {./images/} }

\title{Point and interval estimation of the target dose using weighted regression modelling}

\author{
 Saswati Saha\\
TAGC, Theories and Approaches of Genomic Complexity, \\
 Aix-Marseille Universit\'e, Inserm,\\
Marseille, France.\\
  \texttt{saswati.saha@inserm.fr} \\
   \And
 Werner Brannath\\
FB03 Mathematics / Computer Science - Institute for Statistics, Universit\"at,\\ Bremen, Bremen 28359, Germany. \\
  \texttt{brannath@uni-bremen.de} \\
}
\begin{document}
\maketitle
\begin{abstract}
In a Phase II dose-finding study with a placebo control, a new drug with several dose levels is compared with a placebo to test for the effectiveness of the new drug. The main focus of such studies often lies in the characterization of the dose-response relationship followed by the estimation of a target dose that leads to a clinically relevant effect over the placebo. This target dose is known as the minimum effective dose (MED) in a drug development study. Several approaches exist that combine multiple comparison procedures with modeling techniques to efficiently estimate the dose-response model and thereafter select the target dose. Despite the flexibility of the existing approaches, they cannot completely address the model uncertainty in the model-selection step and may lead to target dose estimates that are biased. In this article, we propose two new MED estimation approaches based on weighted regression modeling that are robust against deviations from the dose-response model assumptions. These approaches are compared with existing approaches with regard to their accuracy in point and interval estimation of the MED. We illustrate by a simulation study that by integrating one of the new dose estimation approaches with the existing dose-response profile estimation approaches one can take into account the uncertainty of the model selection step.  
\end{abstract}


\section{Introduction}
\label{Introduction}
Proper understanding and characterization of the dose-response relationship is a fundamental step in the clinical drug development process. A clear understanding of the dose-response relationship is crucial for addressing two primary questions in a drug development study: 1) Is there an overall effect in the drug, which is typically measured by a clinical endpoint of interest? This is often termed as the Proof of Concept (PoC) step. (ii) If yes, which dose should be selected for further development? There exist an extensive literature \cite{tamhane1996multiple,ruberg1989contrasts,saha2018comparison} addressing the PoC step in dose-response studies. Our focus is more on the target dose estimation that can be reliably used in the later stages of drug development. In Phase II clinical trials, often the aim is to obtain the target dose that produces a clinically relevant effect over the placebo. This is known as the minimum effective dose (MED). We focus mainly on MED estimation in this article. 

Two common statistical approaches that address the above questions are 1) Multiple comparison procedures (MCP), 2) Modeling techniques. Under a MCP approach, the new drug with several active levels is compared with placebo, and PoC is usually established if at least one of the dose is more effective than the placebo control \cite{hochberg1987multiple}. Once the PoC is established, the minimum active dose which gives a statistically significant and clinically relevant effect compared to the placebo is selected as the minimum effective dose (MED) \cite{ruberg1995dose1,tamhane1996multiple,tamhane2002multiple}. Such procedures treat the dose as a qualitative parameter and make very few, if any, assumptions about the underlying dose-response model. A drawback of these procedures is that they cannot extrapolate information beyond the observed dose levels and thereby target dose estimation and inference are restricted to the observed dose-levels under investigations. Modeling techniques, on the other hand, establishes the PoC by first fitting a parametric dose-response model and then comparing it with the ANOVA model with no dose-response effect. If the PoC is established then the underlying regression is utilized for MED estimation \cite{pinheiro2006analysis,saha2018comparison}. However, the true underlying dose-response shape is not known before the start of the trial, and specifying one parametric model for dose-response characterization using modeling techniques often leads to inaccurate and biased inferences.

Bretz et al., \cite{bretz2005combining} proposed an integrated approach namely MCP-Mod (Multiple Comparison Procedures-Modeling) that combines multiple comparisons and modeling techniques to address the above goals in clinical drug development. The MCP-Mod approach begins with a set of potential dose-response models, tests for a significant dose-response effect (PoC) using multiple linear contrasts tests and selects the ``best" model using some model selection criteria among those with a significant contrast test. The ``best" model is then utilized for target dose estimation using regression techniques. A major disadvantage of the MCP-Mod approach is that it makes prior assumptions about the values of the model parameters used in the candidate set which may further lead to a loss in power and unreliable model selection. Several variations of MCP-Mod approaches \cite{dette2015dose,gutjahr2017likelihood,baayen2015testing} are also explored over the past few years which address the apriori parameter value selection issue in the MCP-Mod approach by replacing the multiple contrast test with the log-likelihood ratio tests. A brief comparison study on the above approaches \cite{saha2018comparison} shows that all the aforementioned approaches can establish the PoC with strong control of FWER. Note that the above approaches only differ in the PoC testing step. The model selection and dose selection steps are similar. Even in the MCP-Mod method, the parameter values are re-estimated in the model selection step even though they are pre-specified for the contrast test in the PoC testing step. The above approaches are quite flexible and can address the model uncertainty to some extent. However, it often happens that they lead to the selection of an incorrect model. This model may be adequate for an approximate estimation of the dose-response shape but can be more misleading when estimating the target dose. Moreover, the true dose-response model may not at all belong to the candidate set from which the model is selected.

In this article, two new approaches for estimating MED are proposed, which are robust against model misspecifications. The intention is to fit a weighted regression model where the weights are so designed such that residuals around the target dose are given more importance than the residuals away from the target dose. By doing so, we primarily focus on reducing the bias in estimating the target dose with a misspeciﬁed model at the cost of slightly inﬂating the variance. However, it is important to note that a weighted regression approach is a model-based approach and its' performance will still depend on the choice of the dose-response model.
Hence, it will often be advantageous to apply a dose-response analysis strategy with several parametric candidate models, similar to the ones described in the previous paragraph. By applying one of the following dose-response analysis approaches; the MCP-Mod approach \cite{bretz2005combining} or the approach proposed by Dette et al.,\cite{dette2015dose} or the approach proposed by Baayen et al., \cite{baayen2015testing} to the dataset, one can simultaneously test the PoC (no dose-related effect) and
detect the dose-response shape that best describes the data. When the PoC is rejected, the ‘best’ model can be utilized for target dose estimation using the estimation methods discussed in this article. The intuition lies in the fact that when the model is misclassified, one can leverage the idea of using weights to obtain less biased and thereby better estimates of target doses. 
 
We also propose new confidence intervals for the MED in this article. Confidence intervals contain a wealth of clinically relevant information that is generally not available from P values and usual significance testing \cite{visintainer1998understanding}. Confidence intervals estimation for dose-finding studies have not yet been thoroughly investigated.

An obvious option for interval estimation around the target dose based on modeling techniques is to use the delta method for obtaining asymptotic confidence bounds (Wald confidence limits) around the dose-response curve \cite{bretz2005combining, dette2008optimal, seber1989nonlinear}. This is further discussed later in this article, calling this the classical approach in the rest of the article. The computation of the $100(1-\alpha)\%$ pointwise Wald confidence interval is simple for linear models. But, dose-response relationships are usually best described by models that are non-linear in the parameter as well as the doses. Traditionally, approximate Wald confidence intervals are used for such non-linear models. But often they yield coverage rates that deviate from the nominal level \cite{baayen2015confidence}. Often bootstrap approaches are suggested to improve the coverage probability \cite{bornkamp2009mcpmod, pinheiro2014model}. However, fitting non-linear models are computationally intensive and may result in convergence issues, so that bootstrap intervals based on a large number of model fits can be complex and time-consuming. Baayen and Hougaard \cite{baayen2015confidence} proposed a nice alternative for computing pointwise CI using a profile likelihood approach. In this article, we propose confidence interval estimates for the MED with a more robust weighted estimation approach. We compare the resulting confidence intervals with the classical approach and the two approaches proposed by Baayen and Hougaard \cite{baayen2015confidence}, the percentile bootstrap and the profile likelihood approaches.

The rest of the article is organized as follows: In Section \ref{Parametric set-up}, we will introduce the parametric set-up, in Section \ref{Estimation methods}, we will give a brief overview of the classical approach and give details of the new approaches for MED estimation and inference. Section \ref{Weights} discusses in detail the weight functions used by the simulations presented later in this article and Section \ref{IRWLS} and Section \ref{RobustEstimation} describes the two new MED estimation approaches. Section \ref{Simulation} provides a detailed summary of the findings of our extensive simulation study. The article concludes with a discussion in Section \ref{Discussion}.

\section{Parametric set-up}
\label{Parametric set-up}
A parallel dose-group design is considered here with increasing dose levels $d_0,\, \hdots, \,d_k$ where $d_0$ is the placebo dose. The response $Y$ (which can be an efficacy or a safety variable) is assumed to follow a certain dose-response shape with normally distributed errors:
\begin{equation}
\label{anovamodel}
\begin{aligned}
Y_{ij}= \mu(d_i, \boldsymbol{\theta})+\epsilon_{ij}=\alpha+\beta x_{\boldsymbol{\gamma}}(d_i)+\epsilon_{ij} ,
\\ \quad \epsilon_{ij} \sim N(0,\sigma^2), \quad i=0,\hdots,k,\, j=0,\hdots,n_{i}, 
\end{aligned}
\end{equation}
\noindent where $Y_{ij}$ refers to the response of the $j\textsuperscript{th}$ patient in dose group $i$, $n_i$ is the number of patients in dose group $i$ and $n=\sum_{i=0}^k n_i$ denotes the total sample size. $\mu(d_i, \bm{\theta})$ denotes the non-linear regression function and \newline $\bm{\theta}=\{\alpha,\beta,\bm{\gamma};\alpha,\beta \in \mathbbm{R}, \bm{\gamma}\in \Gamma \subseteq \mathbbm{R}^{p-2}\}$ refers to the vector of corresponding model parameters, where $\alpha$ and $\beta$ are the intercept and slope parameters respectively, and $\boldsymbol{\gamma}$ is the non-linear parameter capturing the non-linear shape of the model, $x_{\boldsymbol{\gamma}}(d)$. $\Gamma$ denote the parameter space for $\boldsymbol{\gamma}$. Table~\ref{tab:table1} shows examples of linear or non-linear
transformations of the dose variable $d$ which can be considered in $x_{\bm{\gamma}}(d)$.
\begin{table}[!htb]
\centering
	\small
	\caption{\textit{Linear or non-linear regression functions considered for the dose-response modelling}}
	\label{tab:table1}
	\scalebox{0.94}{
		\begin{tabular}{p{3cm}p{3.5cm}p{3.5cm}} 
			\hline
			\textbf{ Model } & $\mu(\alpha,\beta,{\gamma},d)$ &$x_{\gamma}(d)$ \\  [0.4ex] \hline 
			\hline
			Linear & $\alpha+\beta d $&$ d $\\ [0.4ex]
			Linear in log-dose &$ \alpha+\beta log(d+c) $&$ log(d+c) $\\  [0.4ex] 
			Emax &    $ \alpha+\beta \frac{d}{ED_{50}+d} $     &    $ \frac{d}{ED_{50}+d}$    \\ [0.7ex] 
			Exponential &     $ \alpha+\beta [exp(\frac{d}{\delta}-1)]$    &  $ [exp(\frac{d}{\delta}-1)]$   \\ [0.7ex] 
			Quadratic & $\alpha+\beta (d+\frac{\beta_2}{\mid\beta\mid}d^2) $&  $d+\frac{\beta_2}{\mid\beta\mid} d^2$ for $\beta_2<0$\\  [0.7ex] 
			Sigmoid \textit{Emax}& $\alpha+\beta \frac{d^h}{ED_{50}+d^h}$ & $\frac{d^h}{ED_{50}+d^h}$\\  [0.7ex] 
			\hline
		\end{tabular}}
		
	\end{table}

\section{Estimation methods}
\label{Estimation methods}
In this section we introduce two new model-based estimation approaches for the minimum effective dose (MED).

The MED is defined as,
$$MED = \inf\{d \in (d_0,d_k] \,| \, \mu(d_i,\boldsymbol{\theta}) > \mu(d_0,\boldsymbol{\theta}) + \Delta\},$$
where $\Delta >0$ denote the clinically relevant effect over the placebo. The classical approach of estimating the MED by modeling techniques is shown in Bretz et al., and Dette et al., \cite{bretz2005combining,dette2008optimal}. An estimator of the MED \cite{bretz2005combining} is given by :
\begin{equation}
\label{BretzMED}
    \widehat{MED} = \inf\{d \in (d_0,d_k]\, | \, \mu(d,\widehat{\boldsymbol{\theta}}) > \mu(d_0,\widehat{\boldsymbol{\theta}}) + \Delta, L_d>\mu(d_0,\widehat{\boldsymbol{\theta}})\},
\end{equation}
where $\widehat{\boldsymbol{\theta}}$ is the least squares estimate of $\boldsymbol{\theta}$ and $L_d$ is the lower $1-2\alpha$ confidence limit of the estimated mean value $\mu(d,\widehat{\boldsymbol{\theta}})$ at dose $d$ based on the model in equation (\ref{anovamodel}). Following Dette et al., \cite{dette2008optimal}, for large sample sizes $\widehat{MED}$ can be approximated as:
 \begin{equation}
 \begin{aligned}
 \label{MED}
  &\widehat{MED}\approx MED(\widehat{\boldsymbol{\theta}}),\\
  &\textit{ where } MED(\boldsymbol{\theta})= x_{\boldsymbol{\gamma}}^{-1}\bigg{(}x_{\boldsymbol{\gamma}}(d_0)+\frac{\Delta}{\beta}\bigg{)}&
  \end{aligned}
 \end{equation}
 with $x_{\boldsymbol{\gamma}}^{-1}$ denoting the inverse of the function $x_{\boldsymbol{\gamma}}$ with respect to the dose variable $d$. 
Asymptotic variance and asymptotic confidence intervals of $\widehat{MED}$ (equation~\ref{BretzMED}) using the classical $\delta$-method \cite{van1988estimating} are derived and detailed in Dette et al., \cite{dette2008optimal}, where the problem of deriving efficient designs for the estimation of MED are investigated. In this article, we use the approximate conﬁdence bounds described in Dette et al., \cite{dette2008optimal} as a benchmark for the conﬁdence intervals developed here. The approximate confidence bounds described by them \cite{dette2008optimal} are referred to as the classical approach in the rest of the article. We want to obtain robust estimates for the MED using modeling techniques that are less biased under a misspecified model. With this intention, we make the following propositions.

In the regression model stated earlier in equation (\ref{anovamodel}), the following sum of squares:
\begin{equation}
\begin{aligned}
\label{SSE}
SSE=\sum_{i,j}(Y_{ij}-\alpha-\beta x_{{\gamma}}(d_i))^2,
\end{aligned}
\end{equation}
are minimized to obtain the least squares estimates. Instead of minimizing the classical $SSE$ we propose to minimize the weighted SSE ($wSSE$),
 \begin{equation}
   \label{weightedSSE}
       wSSE= \sum_{i,j}w_{ij}(d_i,MED(\alpha,\beta,\gamma))(Y_{ij}-\alpha-\beta x_{{\gamma}}(d_i))^2,
   \end{equation}
where the weights depend on the regression estimates, and the MED function given in equation (\ref{MED}). By doing this we are in a way increasing the sample size around the target dose with the hope of getting better estimates of the dose-response curve around the target dose. This can be helpful under model misspecification when the true model is approximated by a similar model from the class of dose-response model families. Hence, if the estimation is more focused on a smaller part of the dose-response curve instead of the entire curve, one can more accurately get estimates around the point of interest.

Weighted least squares regression are generally implemented for locally weighted scatterplot smoothing \cite{cleveland1988locally} or robust estimation methods in heteroscadestic regression models \cite{carroll1982robust}. However, the weights considered in the above approaches are either fixed numbers known apriori or some parametric functions of the predictors that are not related to the regression parameters. The weights in our approach are function of the regression parameters and hence the above approaches cannot be applied to our weighted least squares problem. We consider two estimation approaches in this article. The first approach performs an iterated re-weighted least squares regression (IRWLS), where the solution depends on the optimization algorithm (Newton approach or alternative optimization algorithm). In the second approach, we consider a robust regression (RR) approach \cite{fraiman1983general}, where the weighted least squares estimates are obtained by solving for $\boldsymbol{\theta}$ in the following normal equations:
\begin{equation}
\label{wnormeqn}
 \sum_{i,j}(Y_{ij}-\mu(d_i,\boldsymbol{\theta}))\cdot w_{ij}(d_i,\boldsymbol{\theta})\frac{\partial \mu(d_i,\boldsymbol{\theta})}{\partial \boldsymbol{\theta}}=0.   
\end{equation}
This is analogous to the weighted score regression \cite{antonijevic2010impact} where generalized estimating equations (GEE's) \cite{liang1986longitudinal} are used to solve for the above normal equations. We are instead using 'M'-estimation approach proposed by \cite{huber1967behavior}. This is further elaborated in Section~\ref{RobustEstimation}.

Note that our theory is developed based on the assumption that the data follows a model from the assumed model family. More precisely, with the second approach we show consistency and asymptotic normality of the weighted least squares estimates under the assumption that the model class is correctly specified. We show in Section \ref{Simulation} via simulations that our estimates perform better than the classical least squares estimates if the model family is misspecified.  We further give a heuristic justiﬁcation in Appendix~\ref{Appendix A.1} on how the estimates obtained by solving the above normal equations (\ref{wnormeqn}) can lead to the estimates which minimize the weighted SSE in equation (\ref{weightedSSE}).

\subsection{Weights}
\label{Weights}
\begin{figure}[!htb]
\centering
    \includegraphics{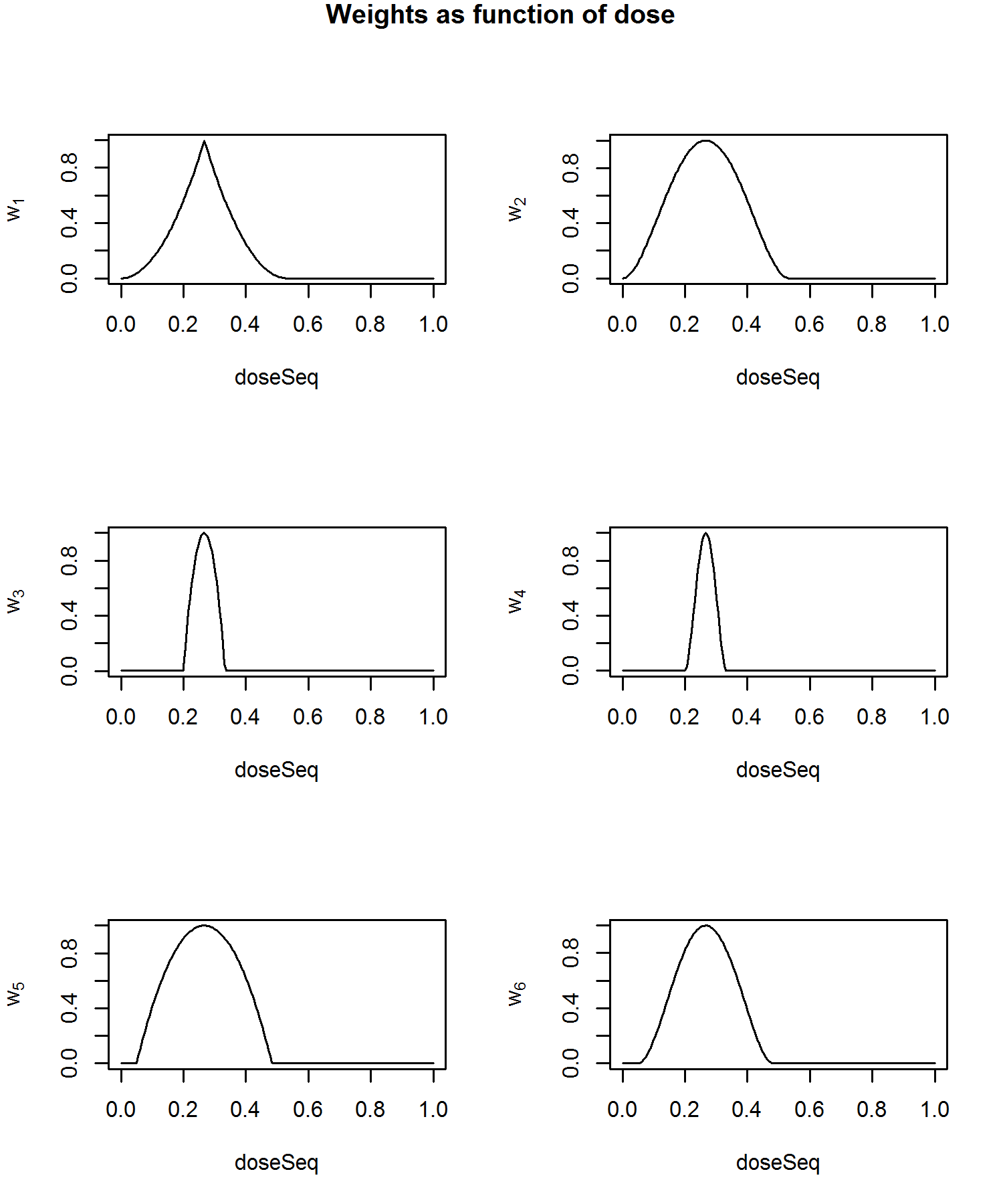}
    \caption{Different weight functions plotted when the underlying true dose-response model is the following emax model: $0.2+0.7\frac{d}{d+0.2}$.The weight functions peak at the MED (0.26 here) and fades away, as one moves away from the MED. This is the MED required to achieve the clinical threshold of $\Delta=0.4$ in the aforementioned emax model.}
    \label{fig:Weights}
\end{figure}

The choice of the weight function is crucial in the above estimation approaches. The aforementioned RR approach makes some asymptotic inference that relies on smooth and continuously differentiable weight functions. This is elaborated in the next section. Even within the class of smooth functions, a wide range of options are available and the choice of the weights is at the discretion of the user. We have considered 6 possible weight functions, some of which are taken from Cleveland and Devlin, \cite{cleveland1979robust,cleveland1988locally} and others are experimental. The six possible weight functions considered in this article are:

\begin{tabular}{p{3cm}p{5cm}}
{\begin{flalign*} 
    w_1(d,d_{MED})& =1-|z|^2 & \\
   w_2(d,d_{MED})& =(1-|z|^2)^2 
 \end{flalign*}}
&
 {\begin{align*}
  \text{where } z=
       \begin{cases}
    \frac{d_{MED}-d}{d_{MED}}     & \quad \text{if } |\frac{d_{MED}-d}{d_{MED}}|<0.9999\\
    0.9999  & \quad \text{o.w}\\
  \end{cases},
  \end{align*}}\\ [-5ex]
{\begin{flalign*} 
    w_3(d,d_{MED})& =1-|z|^2 & \\
   w_4(d,d_{MED})& =(1-|z|^2)^2 
 \end{flalign*}}
&
 {\begin{align*}
  \text{where } z=
       \begin{cases}
    \frac{d_{MED}-d}{\underset{d\in \{d_1,\hdots, d_k\}} \min|d-d_{MED}|}    & \, \text{if } |\frac{d_{MED}-d}{\underset{d\in \{d_1,\hdots, d_k\}} \min|d-d_{MED}|} |<0.9999\\
    0.9999  & \quad \text{o.w}\\
  \end{cases}
  \end{align*}}\\[-5ex]
  {\begin{flalign*} 
    w_5(d,d_{MED})& =1-|z|^2 & \\
   w_6(d,d_{MED})& =(1-|z|^2)^2 
 \end{flalign*}}
&
 {\begin{align*}
  \text{where } z=
       \begin{cases}
    \frac{d_{MED}-d}{\underset{d\in \{d_{(2)},\hdots, d_{(k)}\}} \min|d-d_{MED}|}    & \, \text{if } |\frac{d_{MED}-d}{\underset{d\in \{d_{(2)},\hdots, d_{(k)}\}} \min|d-d_{MED}|} |<0.9999\\
    0.9999  & \quad \text{o.w}.\\
  \end{cases}
  \end{align*}}
\end{tabular}\\
Here $d_{MED}$ is the MED estimate ($MED(\widehat{\boldsymbol{\theta}})$) obtained using equation (\ref{MED}) and $d_{(1)},\hdots, d_{(k)}$ are arranged such that $d_{(1)}$ is closest dose in $\{d_1,\hdots,d_k\}$ to $d_{MED}$ and $d_{(k)}$ is furthest away from $d_{MED}$. The above weight functions are illustrated in Figure \ref{fig:Weights}. The rationale behind considering the weight functions $w_3$ to $w_6$ is to provide a narrow window around the MED where the residuals in $wSSE$ (equation (\ref{weightedSSE})) are given maximum weights.

We have also considered a discrete weight function. The discrete weight function is used only in the IRWLS method because the RR approach is only applicable for continuous weight functions. The discrete weight function considered in this article is:
  \[ w_7(d,d_{MED}) = 
  \begin{cases}
    k_1      & \quad \text{if } d=\underset{d\in \{d_1,\hdots, d_k\}} \argmin|d-d_{MED}|\\
    k_2  & \quad \text{o.w}\\
  \end{cases}
\]
where $k_1$ and $k_2$ are integers with $k_1>k_2$ and $d_{MED}$ is same as the estimated MED ($MED(\widehat{\boldsymbol{\theta}})$) given by equation (\ref{MED}).

\subsection{Iterated re-weighted least squares regression (IRWLS)}
\label{IRWLS}
Weighted non-linear regression using unconstrained optimization is a common problem and a broad variety of standard tools are available. Numerical optimization based on simplex methods like the Nelder-Mead algorithm \cite{nelder1965simplex}, pattern search algorithms \cite{torczon1997convergence} or gradient descent algorithm  can be used to solve the above. However, the weights involved in a non-linear regression problem are typically some linear or non-linear functions that are not dependent on the regression parameters\cite{lim2012accounting}. So far, we have not encountered a situation in literature where the weights in the regression model are treated as functions of the regression parameters. In this article, we propose an iterated re-weighted least squares regression (IRWLS) method by which one can naively obtain estimates for the weighted least squares estimates, where the weights are expressed as a function of the regression parameters. The IRWLS method can be outlined as follows: 
\begin{itemize}[labelindent = 0pt, topsep = 0ex, itemsep=0.5pt]
    \item[Step 1.] Start with the un-weighted non-linear least squares estimates of model equation (\ref{anovamodel}) obtained by minimizing $SSE$ in equation (\ref{SSE}). Denote them by $\widehat{\boldsymbol{\theta}}_0$ and $\hat{\sigma}$.
    \item[Step 2.] The estimates for the subsequent iterations ($i=2,\hdots,100$) can be computed using the follow steps:
    \begin{itemize}[labelindent = 0pt, topsep = 0ex, itemsep=0.5pt]
    \item[a.] The parametric form of the weight functions are known apriori. Examples of such parametric forms are illustrated in Section \ref{Weights}. For the $i^{th}$ iteration, the weights can be obtained based on the estimates of $\widehat{\boldsymbol{\theta}}$ obtained in the $(i-1)^{th}$ iteration. We organize the weights for all the doses in the form of a diagonal matrix $\boldsymbol{W_{\nu \times \nu}}$.
\[
 \boldsymbol{W_{\nu \times \nu}}= \begin{pmatrix}
    w(d_0,d_{MED}) & 0 & \dots & 0  \\
    \vdots & \ddots & \vdots & \vdots \\
    0 & 0 & w(d_,d_{MED}) & 0 \\
    \vdots & \vdots & \ddots & \vdots \\
    0 & 0 & \dots & w(d_k,d_{MED})
  \end{pmatrix},
\]
where $\nu=\sum_i n_i=n$ and $w$  are weight functions, examples of which are provided in Section \ref{Weights} and $$d_{MED} = \inf\{d \in (d_0,d_k] \,| \, \mu(d_i,\boldsymbol{\theta}) > \mu(d_0,\boldsymbol{\theta}) + \Delta\},$$ 
as defined in the previous section. In the subsequent sections we are dropping the ranks of $\boldsymbol{W_{\nu \times \nu}}$ for the sake of convenience.
    \item[b.] Solve $\tau$ for
           $$\boldsymbol{WY}=\boldsymbol{WX}\boldsymbol{\tau},$$
     where $\boldsymbol{\tau}=(\alpha,\beta)$ are the slope and intercept parameters introduced with equation (\ref{anovamodel}) which are updated to the estimate $\boldsymbol{\widehat{\tau}_i}$ by the weighted least squares regression estimates obtained by solving the above. Note that, $X=[1; dose]$ for a linear model and $X=[1; log(dose)]$ for a log linear model. But in a non-linear regression model the design matrix $X$ is a function of the non-linear parameters $\boldsymbol{\gamma}$. An algorithm implemented in the R package \texttt{DoseFinding} \cite{bornkamp18dosefinding} is used here to solve for the non-linear weighted least squares regression. A grid of non-linear parameters $\boldsymbol{\gamma}$ is first selected and for each possible choice of  $\boldsymbol{\gamma}$ in the selected grid, $\alpha$ and $\beta$ estimates are obtained as a solution of the following equation:
      $$ \boldsymbol{WY}=\boldsymbol{WX}_{\boldsymbol{\gamma}}\boldsymbol{\tau} \text{ where } X_{\boldsymbol{\gamma}}=[1;\mu(dose,\boldsymbol{\gamma})]
      $$
     Then the final set of parameters which provide the smallest $SSE$ are selected for the final estimates: $\widehat{\boldsymbol{\theta}_i}=(\widehat{\boldsymbol{\tau_i}},\widehat{\boldsymbol{\gamma_i}})$. $\widehat{\sigma_i}$ is calculated with the new estimate $\widehat{\boldsymbol{\theta}_i}$. 
     \item[c.] Obtain the estimate of $\widehat{MED_i}$ using equation (\ref{MED}) and the expected mean response at the estimated $\widehat{MED_i}$.
       \end{itemize}
     \item[Step 3.] Check for convergence of the regression parameters $\widehat{\boldsymbol{\theta}_i}$ based on one of the following convergence criteria ($tol$ denotes the tolerance limit here):
         \begin{itemize}[wide, labelindent = 0pt, topsep = 0ex, itemsep=0.5pt]
    \item[a.] $\mathrel{\bigg(}\frac{(\widehat{MED}_{new}-\widehat{MED}_{old})}{\widehat{MED}_{old}}\mathrel{\bigg)}^2 \leq tol$
  or if $\widehat{MED} <=0$\;
     \item[b.]  $\mathrel{\bigg(}\frac{(Resp(\widehat{MED}_{new})-Resp(\widehat{MED}_{old}))}{Resp(\widehat{MED}_{old})}\mathrel{\bigg)}^2 \leq tol$
  or if $\widehat{MED} <=0$\;
  \end{itemize}
  where $tol$ (tolerance) is set to $0.001$ (similar to the tolerance limit used in obtaining convergence in the least squares estimates using R package \texttt{nlstools} \cite{baty2015toolbox}). Note that a and b are two alternative options for checking convergence. We conduct simulations firstly with criteria a and then with criteria b. Simulation results in Section \ref{Simulation} are shown with criteria a only, because it performed better than criteria b in our simulations. If the method converges we stop, else we go to Step 4.

   \item[Step 4.] Repeat Step 2-3 till convergence or the maximum iteration is reached. If the algorithm fails to converge or ends in a non-positive MED value, the initial un-weighted non-linear least squares estimates in Step 1 are assigned as the final estimates.
  
\end{itemize}
A more formal description of the algorithm can be found in Appendix~\ref{Appendix A.3}. In the IRWLS approach, elaborated above, the weights are treated as if they are fixed and known before each iteration. Several authors \cite{carroll1982robust,box1974correcting} have proposed robust approaches using iterated least squares algorithm where the first three steps are similar to the algorithm proposed by us. But their objectives were to fit a heteroscedastic model to a regression set-up and propose robust estimates for the parameters under such a set-up. Our objective is different in the sense that we assume a homoscedastic model and we intend to estimate the MED that is robust against model misspecification.
 
 Confidence interval estimates can also be obtained with the above approach. 
 \begin{equation}
    \label{ConfIntIRNLS}
\widehat{MED}_{W}\pm u_{1-a}\frac{\hat{\sigma}_{W}}{\sqrt{n}}b^T(\hat{\alpha}_{W},\hat{\beta}_{W},\boldsymbol{\hat{\gamma}}_{W}){M_{W}}^{-}b(\hat{\alpha}_{W},\hat{\beta}_{W},\boldsymbol{\hat{\gamma}}_{W}).
\end{equation}
 where $\hat{\alpha}_W, \, \hat{\beta}_W, \, \boldsymbol{\hat{\gamma}}_W$ are the weighted estimates from the IRWLS approach, $\widehat{MED}_{W}$ is the MED estimated by the IRWLS approach, $\hat{\sigma}_{W}$ is the corresponding standard deviation obtained with the weighted estimates ($\hat{\alpha}_W, \, \hat{\beta}_W, \text{ and } \boldsymbol{\hat{\gamma}}_W$) and $b(\alpha,\beta,\boldsymbol{\gamma})=\frac{\partial{}}{\partial \boldsymbol{\theta}}MED(\boldsymbol{\theta})$ denotes the gradient of the function $MED$ with respect to $\boldsymbol{\theta}$. ${M_{W}}$ is the information matrix which can be defined as follows:
 $$M_W=\sum_{j=0}^{j=k} w_j(d_j,\theta) g(d_j,\bm{\theta})g^T(d_j,\bm{\theta})\in \mathbbm{R}^{{p}\times{p}}$$
  where the weights $w_j$ are from the weight functions (illustrated in Section~\ref{Weights}) used and \newline $g^{T}(d,\widehat{\boldsymbol{\theta}})=\frac{\partial{\mu(d,\widehat{\boldsymbol{\theta}})}}{\partial \boldsymbol{\theta}}=\bigg{(}1,x_{\hat{\gamma}}(d),\hat{\beta} \frac{\partial{x_{\hat{\gamma}}(d)}}{\partial \gamma_1},\hdots \bigg{)}$ is the gradient of the response function $\mu$ with respect to parameter $\boldsymbol{\theta}$. It should be noted that the above confidence interval can only be obtained when we are considering continuous weight functions and the estimation method is converging.
 Moreover, note that with step 4 in the above algorithm we are safeguarding the method from non-convergence; we propose to use the least squares estimates when we fail to observe any gain in the estimation using the weighted least squares algorithm.

\subsection{Robust regression (RR) }
\label{RobustEstimation}
In this section we will describe a MED estimation approach using robust regression following the works of Ricardo Fraiman \cite{fraiman1983general}.

Huber et al.,\cite{huber1967behavior} considered the class of M-estimates ($\widehat{\boldsymbol{\theta}}_n$) of a parameter $\boldsymbol{\theta}\in \mathbb{R}^k$ given by the solution of the system of equations:
\begin{equation}
\label{equation2}
\sum_{i=1}^n \phi(Z_i,\widehat{\boldsymbol{\theta}}_n)=0
\end{equation}
where $Z_1,\hdots,Z_n \in \mathbb{R}^{q+1}$ are i.i.d variables with common distribution $F$, $\phi:\mathbb{R}^{q+1} \times \mathbb{R}^k \rightarrow \mathbb{R}^k$ is a function such that the expected value $\lambda_F(\boldsymbol{t})=E_F(\phi(Z,\boldsymbol{t}))$ exists for all $\boldsymbol{t}$, and has a unique root at $\boldsymbol{t}=\boldsymbol{\theta_0}$. Huber et al., \cite{huber1967behavior} showed that $\widehat{\boldsymbol{\theta}}_n$ converges to the solution  of the equation:
\begin{equation}
\label{equation3}
\lambda_F(\boldsymbol{t})=E_F(\phi(Z,\boldsymbol{t}))=0
\end{equation}
 Fraiman \cite{fraiman1983general} extends the results shown by Huber et al., \cite{huber1967behavior} for a general class of estimators in a non-linear regression set up:
\begin{equation}
\label{equation1}
Y_i=g(X_i,\boldsymbol{\theta_0})+\epsilon_i  
\end{equation}

where $Z_i=(Y_i,X_i) \in \mathbb{R}^{q+1}$ is a sequence of $ i.i.d.$ random vectors with common distribution $F$,  $X_i \in \mathbb{R}^q$, $Y_i \in \mathbb{R} $, $g:\mathbb{R}^q \times \mathbb{R}^k \rightarrow \mathbb{R}$ is the regression function, $\boldsymbol{\theta_0} \in \mathbb{R}^k$ is the true unknown parameter, $\epsilon_i \,$ \textnormal{is independent of} $X_i$ and has distribution $\mathbbm{N}(0,\sigma^2)$. They consider the same class of estimators given by equation (\ref{equation2}) but where equation (\ref{equation3}) can have more than one solution. They derive robust estimators in a non-linear regression model and their objective was to estimate $\boldsymbol{\theta_0}$ efficiently subject to a bound on the estimator's sensitivity to aberrant data.

We want to apply their method on the dose-response models generalized by equation~(\ref{anovamodel}) but our objective is to derive estimators of $\theta_0$ that will ultimately lead to efficient estimator of MED. In order to describe the M-estimate of $\boldsymbol{\theta_0}$ and their asymptotic distribution, Fraiman \cite{fraiman1983general} made the following propositions and assumptions:\\
Define:
$$P(\boldsymbol{\theta})=\sum_{i=1}^n \phi(Y_i,X_i,\boldsymbol{\theta})$$
Then, the Newton-Raphson (NR) algorithm for (\ref{equation2}) is given by:
\begin{equation}
\label{equation4}
\widehat{\boldsymbol{\theta}}_{n,k+1}=\boldsymbol{\widehat{\theta}}_{n,k}-{\bigg{(}\sum_{i=1}^n \frac{\partial \phi(Y_{i},X_i,\boldsymbol{\widehat{\theta}}_{n,k})}{\partial \boldsymbol{\theta}}\bigg{)}}^{-1} \sum_{i=1}^n \phi(Y_{i},X_i,\boldsymbol{\widehat{\theta}}_{n,k})
\end{equation}
Further define the estimates
 \[ \widehat{\boldsymbol{\theta}}_n =
  \begin{cases}
    \lim_{k\to \infty} \widehat{\boldsymbol{\theta}}_{n,k}      & \quad \text{ if limit exists } \\
   \widehat{\boldsymbol{\theta}}_{n,0}  & \quad \text{o.w}\\
  \end{cases}
\]
where the iteration is started in some consistent initial estimate $\widehat{\boldsymbol{\theta}}_{n,0}$ where $\widehat{\boldsymbol{\theta}}_{n,0} \to \boldsymbol{\theta_0}$.

\paragraph*{Assumptions:}
 \begin{itemize}
 \item[H1:] For each value of $Z=(Y,X), t\mapsto \phi(Y,X,\boldsymbol{t})$ is a continuous function and the set $V_Z=\{\boldsymbol{t} \in \mathbb{R}^k| \phi(Y,X,\boldsymbol{t}) \, \textnormal{is not differentiable}\}$ is a nowhere dense subset of the real line, and if $L$ is a line, the intersection $V\cap L$ is a union of intervals or isolated points.
 \item[H2:] 
  $F$ is continuous, $E_F(\phi(Y,X,\boldsymbol{\theta_0}))=0$, and $E_F(\frac{\partial \phi(Y,X,\boldsymbol{\theta_0})}{\partial \theta})$ is non singular.

  \item[H3:] There exist $d_0>0$ such that
    \begin{equation*}
  \begin{aligned}
    \sup_{\|\boldsymbol{a}-\boldsymbol{\theta_0}\|\leq d_0} \|(\phi(Y,X,\boldsymbol{a})\|\,,  F \textnormal{ is integrable}. \\[-6pt]
    \sup_{\|\boldsymbol{a}-\boldsymbol{\theta_0}\|\leq d_0} \|\frac{\partial (\phi(Y,X,\boldsymbol{a})}{\partial \boldsymbol{\theta}}\| \,, F \textnormal{ is integrable}.
    \end{aligned}
    \end{equation*}
    \item[H4:] $\widehat{\boldsymbol{\theta}}_{n,0}\to \boldsymbol{\theta_0}$ a.s.
    \item[N1:] There exist $d_0>0$ such that
  \begin{equation*}
  \begin{aligned}
    \sup_{\|\boldsymbol{a}-\boldsymbol{\theta_0}\|\leq d_0} \|(\phi(Y,X,\boldsymbol{a})\|^2\,, F \textnormal{ is integrable}. \\[-6pt]
    \sup_{\|\boldsymbol{a}-\boldsymbol{\theta_0}\|\leq d_0} \|\frac{\partial (\phi(Y,X,\boldsymbol{a})}{\partial \boldsymbol{\theta}}\|^2 \,, F \textnormal{ is integrable}.
    \end{aligned}
    \end{equation*}
  \end{itemize}

\begin{theorem}
\label{Theorem1}
Assume H1 to H4. Then we can conclude that
\begin{itemize}
\itemsep-0.5em 
\item[(i)] $\widehat{\boldsymbol{\theta}}_{n}\to \boldsymbol{\theta_0}$ a.s.
\item[(ii)] $\exists \, n_0>0 \text{ such that } \widehat{\boldsymbol{\theta}}_{n}=\lim_{k \to \infty} \widehat{\boldsymbol{\theta}}_{n,k}$ if $n \geq n_o$ a.s. 
\end{itemize}
\end{theorem}
\begin{theorem}
\label{Theorem2}
Assume H1 to H4 and N1.\\
\noindent Let $V=E_F(\phi(Y,X,\boldsymbol{\theta_0})\phi(Y,X,\boldsymbol{\theta_0})^t)$ and 
$A=E_F(\frac{\partial \phi(Y,X,\boldsymbol{\theta_0})}{\partial \boldsymbol{\theta}})$. \\
Then, we have $$ \sqrt{n}(\widehat{\boldsymbol{\theta}}_{n} - \boldsymbol{\theta_0}) \xrightarrow{D} \mathrel{N}(0,A^{-1}V{A^{-1}}^t)$$
\end{theorem}
 In order to apply the above theorem to our situation we substitute $Y_i$ by $Y_{ij}$ and $X_i$ by dose $d_{ij}$ and define:
\begin{equation}
\label{FraimanLoss_True}
    \phi(Y_{ij},d_{ij},\boldsymbol{\theta})=(Y_{ij}-\mu(d_{ij},\boldsymbol{\theta}))\cdot w_{d_{ij}}(d_{ij},\boldsymbol{\theta})\frac{\partial \mu(d_{ij},\boldsymbol{\theta})}{\partial\boldsymbol{\theta}}
\end{equation}
where $Y_{ij}$ are the responses and $d_{ij}(=d_i)$ are the dose levels from the set-up (\ref{anovamodel}). Furthermore, to align with the model framework of Fraiman\cite{fraiman1983general}, it is assumed that the dose is a random variable $D$ and doses are so allocated such that $\frac{n_i}{n} \to p_i,$ where $p_i=P(D=d_i)$ when $n\to \infty \, \forall \, i \in \{0,\hdots,k\}$. The distribution of $\epsilon_{ij}$ is assumed to be independent from the dose. For the loss function in equation (\ref{FraimanLoss_True}) and smooth weight functions in Section~\ref{Weights}, one can easily verify that the assumptions $H1$ to $H4$ and $N1$ are satisfied for the dose-response model in equation (\ref{anovamodel}). Appendix~\ref{Appendix A.2} gives further elaborations of how the above assumptions are verified with the above choice of the loss function in equation (\ref{FraimanLoss_True}) and with smooth weight functions shown in Section \ref{Weights}. 

The ordinary least squares estimates are treated as our initial estimates and then a sequence $\widehat{\boldsymbol{\theta}}_{n}$ are obtained satisfying equation (\ref{equation4}). Further justifications of why the least squares estimates can be used as initial estimates in our method are given in the Appendix \ref{Appendix A.1}. Following Theorem \ref{Theorem1} and Theorem \ref{Theorem2}, the sequence of $\widehat{\boldsymbol{\theta}}_{n} $ is distributed as follows:
\begin{equation}
\label{MED_RR}
    \sqrt{n}(\widehat{\boldsymbol{\theta}}_{n} - \boldsymbol{\theta_0}) \xrightarrow{D} \mathrel{N}(0,A^{-1}V{A^{-1}}^t)
\end{equation}  
where 
\begin{equation} \label{eq1}
\begin{split}
V & = E_F[(Y-\mu(d,\boldsymbol{\theta_0}))^2\cdot w^2(d,\boldsymbol{\theta_0})\cdot h(d,\boldsymbol{\theta_0})h^t(d,\boldsymbol{\theta_0})]\\
 \end{split}
\end{equation}
with $h(d_i,\boldsymbol{\theta})=\frac{\partial \mu(d_i,\boldsymbol{\theta})}{\partial \boldsymbol{\theta}}$ and
\begin{equation} \label{eq2}
\begin{split}
A &=E_F\bigg(\frac{\partial \phi(Y,d,\boldsymbol{\theta_0})}{\partial \boldsymbol{\theta}}\bigg)\\
&= E_F\bigg{(}-w(d,\boldsymbol{\theta_0})\frac{\partial \mu(d,\boldsymbol{\theta_0})}{\partial \boldsymbol{\theta}}{\bigg{(}\frac{\partial \mu(d,\boldsymbol{\theta_0})}{\partial \boldsymbol{\theta}}\bigg{)}}^{t}+(Y-\mu(d,\boldsymbol{\theta_0}))\cdot \frac{\partial[{w(d,\boldsymbol{\theta_0})\frac{\partial \mu(d,\boldsymbol{\theta_0})}{\partial \boldsymbol{\theta}}}]}{\partial \boldsymbol{\theta}}\bigg{)}
\end{split}
\end{equation}
 Finally, the MED estimate with the RR approach can be obtained using the equation (\ref{MED}) where estimate of $\gamma$ and $\beta$ are components of the estimates $\widehat{\boldsymbol{\theta}}$ obtained by solving for equation (\ref{equation3}) using NR algorithm (equation \ref{equation4}). The asymptotic distribution of $\widehat{\boldsymbol{\theta_n}}$ shown in equation~(\ref{MED_RR})  can be approximated by plugging-in the empirical estimates of $V$ and $A$.
 
 Furthermore, the asymptotic distribution of $\widehat{MED}$ can also be derived by applying the delta method to equation (\ref{MED_RR}) similar to Dette et al., \cite{dette2008optimal}:
\begin{equation}
\label{Var_MED_RR}
    \sqrt{n}(MED(\widehat{\boldsymbol{\theta}}_{n}) - MED(\boldsymbol{\theta_0}))\xrightarrow{\mathbb{D}} \mathrel{N}(0,\nabla MED(\boldsymbol{\theta_0})^{'}\cdot(A^{-1}V{A^{-1}}^{t})\cdot \nabla MED(\boldsymbol{\theta_0}))
\end{equation} 
where $\nabla MED(\boldsymbol{\theta})$ is the gradient of the function
$MED(\boldsymbol{\theta})$ (equation \ref{MED}). We can use the empirical estimate of the above variance to obtain the conﬁdence intervals for MED with the RR approach.

\begin{figure}[!htb]
    \centering
    \includegraphics[scale=0.5]{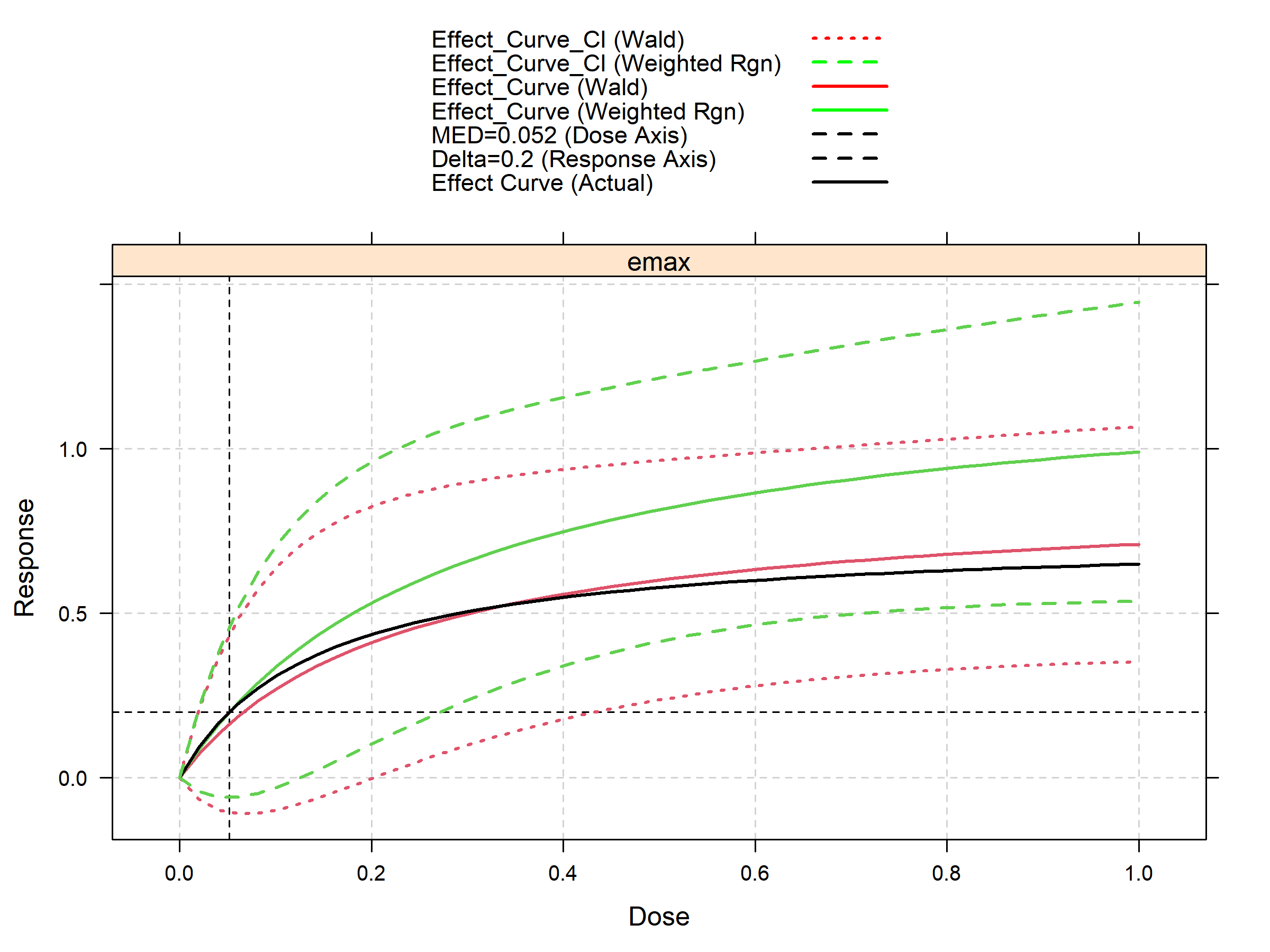}
   \caption{Fitted Emax model on a data simulated from the Emax model in Table \ref{tab:SimulationTableEC} using both the classical approach and RR approach (with weight $w_5$). The solid red line labeled as \texttt{Effect\_Curve} (Wald) denote the effect curve and the dotted black curve labeled as \texttt{Effect\_Curve\_CI} (Wald) denote the confidence bounds for the effect curve using the classical approach and the solid blue line labeled as \texttt{Effect\_Curve} (Weighted Rgn) and the dotted green curve labeled as \texttt{Effect\_Curve\_CI} (Weighted Rgn) denote corresponding effect curve and confidence interval for the effect curve  using the RR approach. The black solid line labeled as \texttt{Effect Curve} (Actual) is the effect curve of the emax model from which the data is simulated. The desired threshold $\Delta=0.2$ needed to obtain the MED is shown by a dotted vertical black line and the actual $MED=0.052$ is indicated by a dotted horizontal black line.}
    \label{fig:EC_Illus}
\end{figure}

To better understand the advantages of using the proposed estimation approaches in this article, we present here a graphical illustration in Figure \ref{fig:EC_Illus}. We have simulated a dataset from an emax model with parameters shown in Table \ref{tab:SimulationTableEC} and then plotted the effect curve, the curve that models the response curve with the placebo response subtracted from it. The effect-curve is estimated with the classical approach and the RR approach (with weight $w_5$) in Figure \ref{fig:EC_Illus}. Since we are interested in the minimum effective dose, the effect curve makes more sense than the full model curve, which models the observed response. Confidence intervals for the effect curve are also plotted in Figure \ref{fig:EC_Illus}. For the desired effect of $\Delta=0.2$, the minimum effective dose is $0.052$. We observe that the effect curve from the classical, unweighted least squares regression shows a uniform deviation from the true curve across the full dose range, however, is less precise than the RR method in the estimation for the effect close to the true MED. Furthermore, the confidence interval of the effect curve shrinks for the RR method (\texttt{Effect\_Curve\_CI} (Weighted Rgn)) as compared to the classical approach effect curve (\texttt{Effect\_Curve\_CI} (Wald)) around the actual MED and widens as the dose increases away from the MED. Thus our objective of getting more accurate dose-response estimates around the target dose is fulfilled in the illustrated data with the weighted regression RR approach.

In the following section, we illustrate the implementation of the above methods and compare their performance with the help of simulations.

\section{Simulation Study}
\label{Simulation}
An extensive simulation study is conducted to evaluate the performance of the new methods proposed in this article and benchmark them against existing approaches. To evaluate the estimation accuracy of the MED, we have compared the MED estimates from the RR and IRWLS approach with the classical MED estimates proposed by Bretz et al., \cite{bretz2005combining} (in equation (\ref{BretzMED})). As already stated in Section~\ref{Introduction}, the new estimation approach is proposed to get MED estimates that are robust not only under the ‘true’ model assumptions but also under model misspecification. We have conducted two sets of simulations to evaluate the MED estimation accuracy. In the first set, we have simulated data from two popular data-generating shapes in clinical trials, the Emax and sigmoidal Emax models, and estimated the MED both under the true model and a misspecified model. The choice of the misspecified model is elaborated later in this section. 

In a second set of simulations, we generate data from several data-generating shapes, apply the MCPMod method to simultaneously test the PoC (no dose-related effect) and detect the dose-response shape that best describes the data. If the PoC is rejected, the 'best' model is utilized for MED estimation using both the classical MED estimates proposed by Bretz et al., \cite{bretz2005combining} and the RR approach. MCPMod is considered here instead of other dose-response strategies \cite{gutjahr2017likelihood,baayen2015testing,dette2015dose} because it was observed in Saha and Brannath, \cite{saha2018comparison} that the model selection performance of all the dose-response strategies were similar and the implementation of MCPMod is faster than the other strategies. Since our primary motivation is to enhance the dose-finding step of the existing dose-response analysis strategies \cite{saha2018comparison}, we want to evaluate the performance of our approach when integrated with the MCPMod approach. RR approach integrated with MCPMod approach is referred to as the MCPMod-RR approach in the rest of the article. The original MED estimates used in Bretz et al., \cite{bretz2005combining} is referred to as the classical MCPMod approach in the rest of the article.  In the previous sections, we have also proposed conﬁdence interval estimates for the MED with the RR approach and the IRWLS approach. In this section, after evaluating the MED estimation accuracy, we compare our conﬁdence intervals with the classical approach \cite{dette2008optimal} and the two approaches proposed by Baayen and Hougaard, \cite{baayen2015confidence}, the percentile bootstrap and proﬁle likelihood approaches. The coverage of the $95\%$ confidence interval for the MED estimates is compared both under the true model and the misspecified model for the aforementioned methods.

The study design used for the simulations is similar to the case study mentioned in Bretz et al., \cite{bretz2005combining}. We investigate five dose levels $(d = 0,\, 0.05,\, 0.2,\, 0.6,\, 1)$, with a single endpoint measured per patient, $Y_i\sim \mathcal{N}(\mu(d), \sigma^2)$. A standard deviation of $\sigma=0.65$ is used in our simulations. This is consistent with the estimated residual SD observed for the case study \cite{bretz2005combining}. MED is estimated corresponding to $\Delta=0.4$ in both the scenarios. The following section describes the two simulation scenarios for evaluating the accuracy of the different MED estimation approaches:

\paragraph*{Simulation Scenario 1:}  Two different data generating
dose-response shapes are
investigated for the mean responses $\mu$. They are defined in Table \ref{tab:SimulationTable}. They have the property that at $d = 0$ the response
value is about $0.2$ and the maximum response is about $0.8$ i.e., maximum dose effect of about $0.6$. For each dose-response shape in Table \ref{tab:SimulationTable}, $5000$ simulated datasets were generated. For each dose-response shape in Table \ref{tab:SimulationTable}, we first fit the true model using the classical and new approaches. We apply our methods with all weights mentioned in Section \ref{Weights} to check which weights are performing well. After analyzing the true model fit we analyze the methods under model misspecifications. For data generated from the sigmoidal Emax model in Table \ref{tab:SimulationTable}, we fit the data also with a linear and an emax model using both, the classical and RR approach. The IRWLS approach is not considered in these simulations because it is not giving good results if the model is not correctly specified. The linear model is chosen because it was observed in Saha and Brannath, \cite{saha2018comparison} that the sigmoidal Emax model was often misclassified as linear model in the simulation results, especially for small sample sizes. We have considered the emax model as it is a commonly used dose-response model and generally considered in the candidate set of the dose-response estimation methods\cite{bretz2005combining, dette2015dose, baayen2015testing, gutjahr2017likelihood}. We analyze the performance of the RR approach under the true model for sample size per dose group: $n=25$ and under model mis-specifications for two sample sizes per dose group: $n=25$ and $n=50$. 

\begin{table}[htb!]
\centering

		 			 	\tabcolsep4pt
		 			 	\renewcommand{\arraystretch}{1.0} 
		 			 	\caption{ Data generating dose-response shapes}
	\label{tab:SimulationTable}
		 			 	\begin{tabular}{p{2cm}p{4cm}p{4cm}}
		 			 		\hline\hline
		 			 		
		 			 		Model  &  Simulated from & Fitted on  \\
		 			 		\hline

		 			 		\multicolumn{1}{l}{{Emax}} & {$0.2+ 0.7 \frac{d}{0.2+d}$} &emax \\ 
		 			 				 		\multicolumn{1}{l}{Sigmoidal} & $0.2+ 0.615 \frac{d^{4}}{0.4^{4}+d^{4}}$& sigEmax, linear, emax\\ 
		 			 				 		
		 			 		\hline\noalign{\smallskip}
		 			 	\end{tabular}
		 			 	
		 			 \end{table}

\paragraph*{Simulation Scenario 2:} In the second simulation scenario, we have used a similar design as in the first but the data are generated from seven non-linear models as tabulated in Table~\ref{tab:SimulationTableMM-RR}. Since the IRWLS approach did not perform as good as the RR approach in the earlier simulations, it is not considered in the second simulation scenario. We apply our methods with selected weights, $w_4$, $w_5$ and $w_6$ from Section \ref{Weights} because the above weights show consistent results in Simulation Scenario 1. Note that results are shown only for weight $w_6$ in Section~\ref{SimulationResults} as the other weights are giving comparable results. We have also shown the impact on the MED estimation approaches when the data-generating shapes included in the candidate set of MCPMod deviate somewhat from the models in Table~\ref{tab:SimulationTableMM-RR}.  To achieve this, we have added some noise to the parameters of the linear, emax and sigmoidal Emax model. The new set of data generating shapes with added noise are illustrated in Table~\ref{tab:SimulationTable_newMM-RR}. We have introduced two noise variables, $\epsilon_1$ and $\epsilon_2$ in Table~\ref{tab:SimulationTable_newMM-RR} that follows uniform distribution with ranges $(0, 0.01)$ and $(0, 0.06)$, respectively.
To summarize, we have applied the classical MCPMod approach and the MCPMod-RR approach to the data generated from both Table~\ref{tab:SimulationTableMM-RR} and Table~\ref{tab:SimulationTable_newMM-RR}. The results are shown in Section~\ref{SimulationResults}.

\begin{table}[!htb]
\parbox{.45\linewidth}{
\centering
\tabcolsep3pt
		 			 			\caption{\textit{ Data generating dose-response shapes}}
		 			 	\renewcommand{\arraystretch}{1.3} 
		 			 		\begin{tabular}{p{2cm}p{5cm}}
		 			 		\hline\hline
		 			 		Model  &  $\mu_l(\alpha,\beta,{\gamma}_l,d)$  \\
		 			 		\hline
		 			 		\multicolumn{1}{l}{Linear} &  {\textcolor{blue}{$0.2+0.6d$}}  \\ 
		 			 		\multicolumn{1}{l}{Linlog} &{ $0.74+ 0.33{log(d+0.2)}$} \\ [0.6ex]
		 			 		\multicolumn{1}{l}{Emax1} & {\textcolor{blue}{$0.2+ 0.7 \frac{d}{0.2+d}$}} \\ [0.7ex]
		 			 		\multicolumn{1}{l}{{Emax2}} & {$0.1+ 0.3 \frac{d}{0.01+d}$} \\ [0.7ex]
		 			 		\multicolumn{1}{l}{Sigmoidal Emax} & \textcolor{blue}{$0.2+ 0.6 \frac{d^{4}}{0.4^{4}+d^{4}}$}\\ [0.7ex]
		 			 		\multicolumn{1}{l}{Power} & $0.2+ 0.6 d^{0.5}$\\ [0.6ex]
		 			 		\multicolumn{1}{l}{Truncated Logistic} & $0.2+\frac{0.682}{(1+exp(10(0.8-d)))}$\\ [0.7ex]
		 			  		
		 			 		\hline\noalign{\smallskip}
		 			 	\end{tabular}
		 		 	\label{tab:SimulationTableMM-RR}

}
\hfill
\parbox{.45\linewidth}{
\centering
\tabcolsep3pt
		 			 			\caption{\textit{ Data generating dose-response shapes}}
		 			 	\renewcommand{\arraystretch}{1.3} 
		 			 		\begin{tabular}{p{2cm}p{5cm}}
		 			 		\hline\hline
		 			 		Model  &  $\mu_l(\alpha,\beta,{\gamma}_l,d)$  \\
		 			 		\hline
		 			 		\multicolumn{1}{l}{Linear} &  {\textcolor{blue}{$0.2+(0.6+\epsilon_2)d$}}  \\ 
		 			 		\multicolumn{1}{l}{Linlog} &{ $0.74+ 0.33{log(d+0.2)}$} \\ [0.6ex]
		 			 		\multicolumn{1}{l}{Emax1} & {\textcolor{blue}{$0.2+ (0.7+\epsilon_1) \frac{d}{0.2+\epsilon_1+d}$}} \\ [0.7ex]
		 			 		\multicolumn{1}{l}{{Emax2}} & {$0.1+ 0.3 \frac{d}{0.01+d}$} \\ [0.7ex]
		 			 		\multicolumn{1}{l}{Sigmoidal Emax} & \textcolor{blue}{$0.2+ (0.6+\epsilon_1) \frac{d^{4}}{(0.4+\epsilon_1)^{4}+d^{4}}$}\\ [0.7ex]
		 			 		\multicolumn{1}{l}{Power} & $0.2+ (0.6+\epsilon_2) d^{0.5+\epsilon_1}$\\ [0.6ex]
		 			 		\multicolumn{1}{l}{Truncated Logistic} & $0.2+\frac{0.682}{(1+exp(10(0.8-d)))}$\\ [0.7ex]
		 			  		
		 			 		\hline\noalign{\smallskip}
		 			 	\end{tabular}
		 		 	\label{tab:SimulationTable_newMM-RR}

}
\end{table}
		 			  
In order to evaluate the coverage of the new confidence bounds for the MED, we conduct a simulation study similar to the ones mentioned in Baayen and Hougaard, \cite{baayen2015confidence}. Following them, we simulate data for the same five dose levels ($0,\, 0.05,\, 0.2,\, 0.6,\, 1$) from the Emax and sigEmax models, shown in Table \ref{tab:SimulationTableEC}.

\begin{table}[htb!]
\centering

		 			 	\tabcolsep4pt
		 			 	\renewcommand{\arraystretch}{1.0} 
		 			 	\caption{ Data generating dose-response shapes for coverage probability evaluation}
	\label{tab:SimulationTableEC}
		 			 	\begin{tabular}{p{2cm}p{5cm}}
		 			 		\hline\hline
	                  Model  &  Simulated from \\
		 			 		\hline
              Emax & {$0.32+ 0.74\frac{d}{0.14+d}$}\\ 
		 	  Sigmoidal & $0.32+ 0.66\frac{d^{4}}{0.3^{4}+d^{4}}$\\ 
	\hline\noalign{\smallskip}
		 			 	\end{tabular}
		          \end{table}		 
		          
The standard deviation of $\sigma=0.65$ is used like in the earlier scenarios. For each dose-response shape in Table \ref{tab:SimulationTableEC}, $2000$ simulations are considered. 

The coverage of the $95\%$ confidence interval for the MED (corresponding to $\Delta=0.3$) are evaluated, both under the true model and misspecified model, for the following methods:  confidence interval using a) the classical approach \cite{dette2008optimal}, b) the bootstrap approach, c) the profile likelihood approach introduced by Baayen and Hougaard, \cite{baayen2015confidence} and d) the weighted regression approach with the RR and IRWLS method introduced in this article. A different scenario from earlier is considered in Table \ref{tab:SimulationTableEC} for the evaluation of the coverage probability because we wanted to keep the simulation set-up the same as the simulation set-up used in Baayen and Hougaard, \cite{baayen2015confidence}. The bootstrap approach for confidence intervals can be parametrical or non-parametrical. Under the parametric bootstrap approach, maximum likelihood estimates from the dose-response model are first obtained and then samples are generated under the model assumptions in (equation \ref{anovamodel}) using these maximum likelihood estimates. The MED is then obtained from these bootstrap samples using the estimates proposed by Bretz et al., \cite{bretz2005combining} in equation (\ref{BretzMED}). The lower and upper $100(1-\alpha)\%$ quantiles of the MED obtained from the bootstrapped samples are used for the confidence intervals of the MED. However, the above bootstrap approach did not perform well in our simulation scenarios. So we used the percentile bootstrap approach suggested in Baayen and Hougaard, \cite{baayen2015confidence}. It can be outlined as follows: Select a grid of doses $d_1,\,\hdots,\,d_G$ and for the selected grid, the aim is to obtain pointwise confidence intervals for the expected outcomes $\hat{y}=\mu(d_g,\widehat{\boldsymbol{\theta}})$ based on a certain dose-response model $\mu$ (equation \ref{Parametric set-up}). For each dose group $i$, $n_i$ samples are drawn with replacement from the observed data. This is repeated $B$ times to generate $B$ bootstrap samples ($y_b^*,\, b=\{1,\hdots,B\}$). For each bootstrap sample $b$: 1) fit the dose-response model parameters with the bootstrap data and the given dose-response model and 2) compute the expected outcomes at pre-selected grid of doses ($\widehat{y}_b^*(d_g),  g=1,\hdots,G$). For each bootstrap sample, $100(1-\alpha)\%$ pointwise confidence intervals for the expected outcomes $(\widehat{y}_{(\alpha)}^*(d_g),\widehat{y}_{(1-\alpha)}^*(d_g))$ are computed for all doses. Finally, inverse regression techniques are applied to the above confidence bounds to obtain the confidence intervals for the MED. We use the above percentile bootstrap approach to benchmark the performance of our approach with regard to the coverage probability. 

Baayen and Hougaard, \cite{baayen2015confidence} also suggested another approach where one first reformulate the regression model in equation (\ref{anovamodel}) in terms of the difference of the expected response to the placebo, i.e. $\beta^*=y^*-\theta_0$ at a specific known dose $d^*$ $>0$, where $y^*$ is the expected response at $d^*$. Then they showed how to obtain profile likelihood confidence intervals for the parameter $\beta^*$. Followed by this, they select a grid of increasing doses $d_1,\,\hdots,\,d_G$, similar to the earlier approach, and for each dose $d_g \text{ where } g \in \{1,\hdots,G\}$ obtain the profile likelihood confidence interval of $\beta_g$, i.e, the corresponding value of $\beta^*$ at dose $d^*=d_g$. This leads to the confidence interval for the effect curve using a profile likelihood approach for the selected grid of doses. To obtain an estimate of the MED, they proposed to take the smallest dose amongst the pre-selected grid,
which achieves an effect of $\Delta$ over the placebo. To obtain a
confidence interval for the MED using the profile likelihood approach they proposed to take the inverse of the confidence bounds of the effect curve at response value $\Delta$. More details can be found in their article \cite{baayen2015confidence}.

\section{Simulation Results}
\label{SimulationResults}
In this section, we present the results of the simulation studies described above and provide a detailed discussion on the pros and cons of the different MED estimation methods mentioned in this article. 

The boxplot distribution of the estimated MED under the different methods and different scenarios gives an idea of the MED estimation accuracy. Additionally, a table is added to each boxplot with measures for the distance of the estimated MED to the true MED (MED: the minimum dose producing an improvement of $\Delta = 0.4$
over the placebo). Since the true MED changes with the
model, the performance of the estimator $\widehat{MED}_i$ is measured in terms of its relative deviation $R_i$ from the true MED, where $R_i= \frac{100(\widehat{MED}_i - MED)}{MED}$. The mean, median and interquartile range (IQR) of $R_i$ from $5000$ simulation runs characterize the relative bias and variability of
$\widehat{MED}_i$.

\begin{figure}%
\centering
\subfigure[MED estimation for data simulated from the Emax Model and fitted with the Emax model]{
\label{fig:Emax}
\includegraphics[scale=0.95]{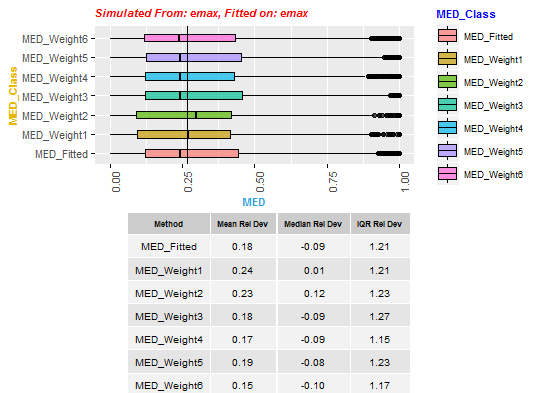}}
\subfigure[MED estimation for data simulated from the sigEmax Model and fitted with the sigEmax model]{%
\label{fig:sigEmax}%
\includegraphics[scale=0.95]{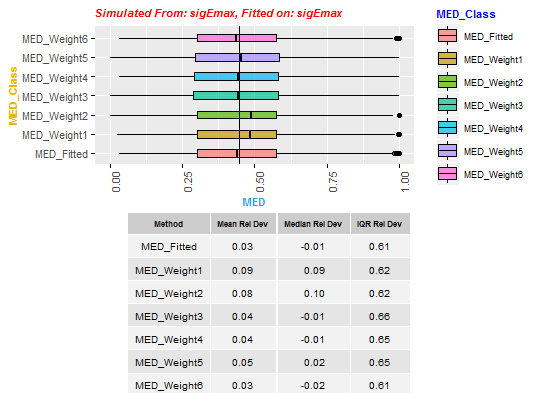}}%
\caption[Short caption]{Boxplot distribution of MED estimated under the classical and RR approach for the different weights given in Figure \ref{fig:Weights}. This is followed by a table which summarizes the mean, median and IQR of $R_i$ for the 5000 simulation trials. Figure (a) and Figure (b) show the MED estimated with the true model for the data simulated from the emax model and sigEmax model in Table \ref{tab:SimulationTable}, respectively. \texttt{MED\_Fitted} denotes the MED estimated using the classical approach and \texttt{MED\_Weights1},$\cdots$,\texttt{MED\_Weights6} denote the MED estimated using RR approach under then different choice of weights $w_1,\cdots, w_6$ given in Figure \ref{fig:Weights}.}
 \label{fig:RR}
\end{figure}

\begin{figure}%
 \centering
\subfigure[MED estimation for data simulated from the Emax Model and fitted with the Emax model]{%
\label{fig:Emax2}%
\includegraphics[scale=0.76]{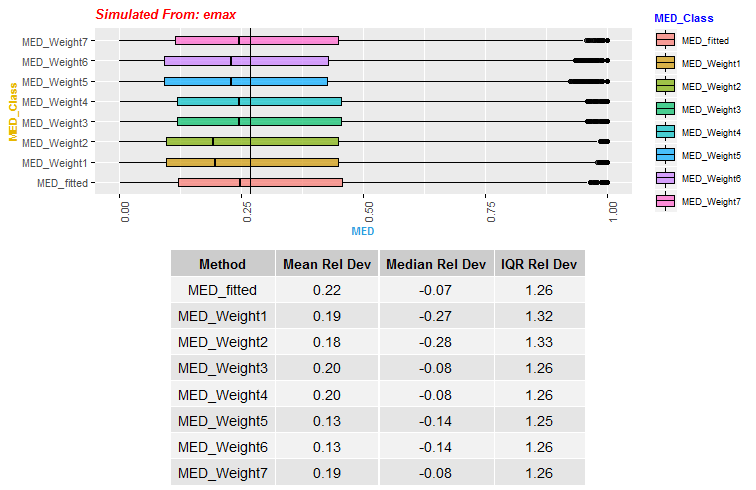}}
\qquad
\subfigure[MED estimation for data simulated from the sigEmax Model and fitted with the sigEmax model]{%
\label{fig:sigEmax2}%
\includegraphics[scale=0.76]{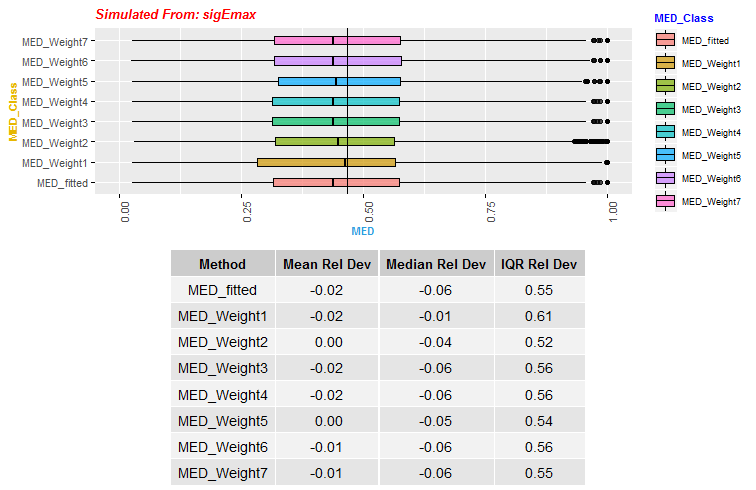}}
\caption{Boxplot distribution of MED estimated under the classical and IRWLS approach for the different weights given in Figure \ref{fig:Weights}. This is followed by a table which summarizes the mean, median and IQR of $R_i$ for the 5000 simulation trials. Figure (a) and Figure (b) show the MED estimated with the true model for the data simulated from the emax model and sigEmax model in Table \ref{tab:SimulationTable}, respectively. \texttt{MED\_Fitted} denotes the MED estimated using the classical approach and \texttt{MED\_Weights1}$,\cdots, $\texttt{MED\_Weights6}, \texttt{MED\_Weights7} denote the MED estimated using IRWLS approach under then different choice of weights $w_1, \cdots, w_7$ given in Figure \ref{fig:Weights}.}
 \label{fig:IRNLS}
\end{figure}

\begin{figure}%
 \centering
\subfigure[MED estimation for data simulated from the sigEmax model and fitted with the linear and sigEmax model for sample size 25]{%
\label{fig:sigEmaxlinear1}%
\includegraphics[scale=0.73]{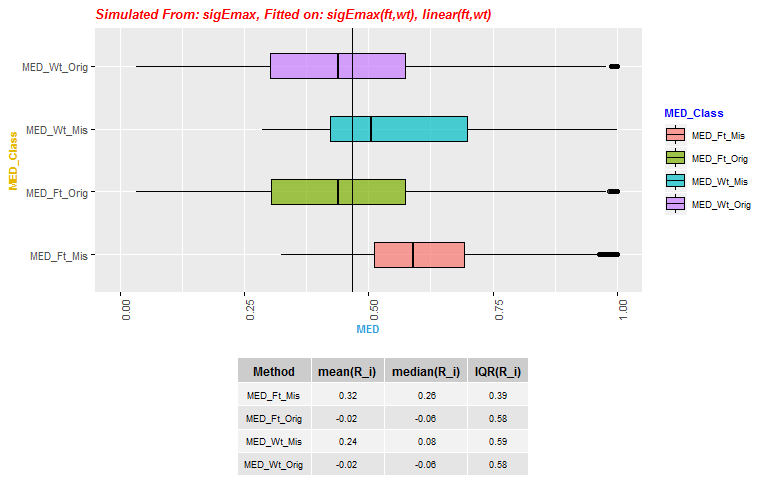}}
\qquad
\subfigure[MED estimation for data simulated from the sigEmax model and fitted with the linear and sigEmax model for sample size 50]{%
\label{fig:sigEmaxlinear2}%
\includegraphics[scale=0.73]{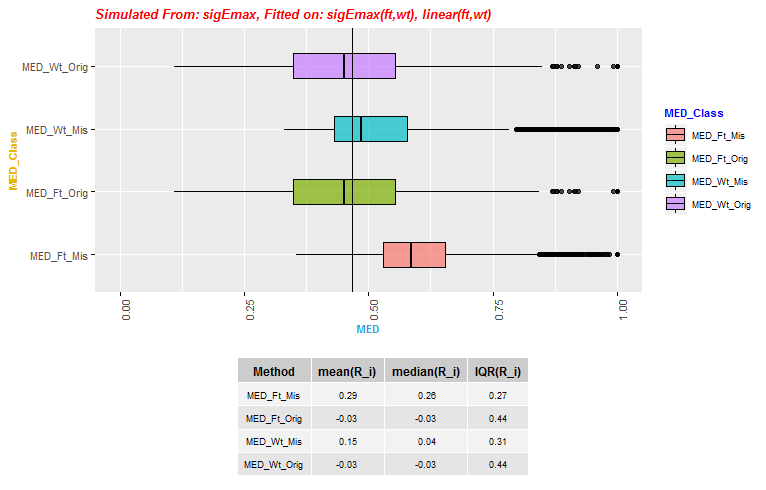}}
\caption{Boxplot distribution of MED estimated under the classical and RR approach for weight $w_6$. This is followed by a table which summarizes the mean, median and IQR of $R_i$ for the 5000 simulation trials. Figure (a) and Figure (b) show the MED estimated with the true model (sigEmax) and the mis-specified model (linear) for the data simulated from the sigEmax model in Table \ref{tab:SimulationTable} with sample size $25$ and $50$, respectively. \texttt{MED\_Ft\_Orig} and \texttt{MED\_Ft\_Mis} denote the MED estimated using the classical approach with the true and mis-specified model, respectively. \texttt{MED\_Wt\_Orig} and \texttt{MED\_Wt\_Mis} denote the MED estimated using the RR approach with the true and mis-specified model, respectively.}
 \label{fig:RRsigEmaxLin}
\end{figure}

\begin{figure}%
 \centering
\subfigure[MED estimation for data simulated from the sigEmax model and fitted with the emax and sigEmax model for sample size 25]{%
\label{fig:sigEmaxEmax1}%
\includegraphics[scale=0.73]{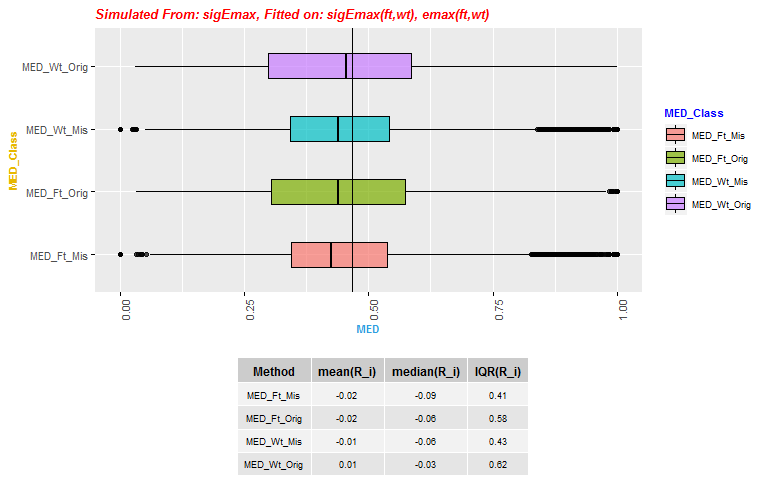}}
\qquad
\subfigure[MED estimation for data simulated from the sigEmax model and fitted with the emax and sigEmax model for sample size 50]{%
\label{fig:sigEmaxEmax2}%
\includegraphics[scale=0.73]{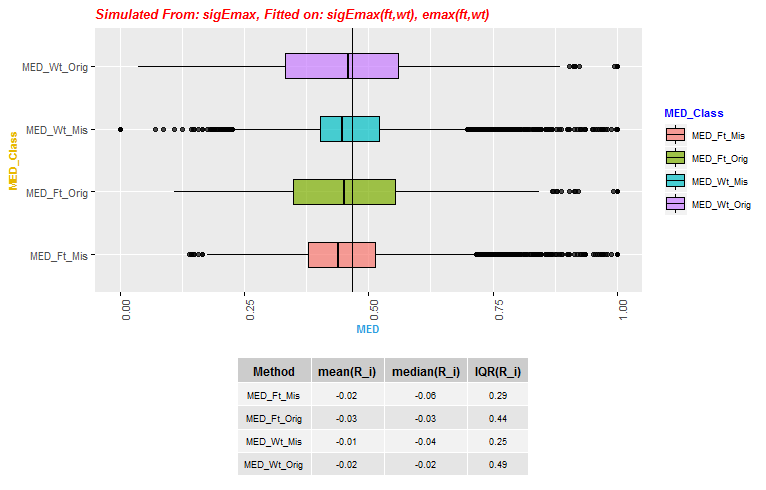}}
\caption{Boxplot distribution of MED estimated under the classical and RR approach for weight $w_5$. This is followed by a table which summarizes the mean, median and IQR of $R_i$ for the 5000 simulation trials. Figure (a) and Figure (b) show the MED estimated with the true model (sigEmax) and the mis-specified model (emax) for the data simulated from the sigEmax model in Table \ref{tab:SimulationTable} with sample size $25$ and $50$, respectively. \texttt{MED\_Ft\_Orig} and \texttt{MED\_Ft\_Mis} denote the MED estimated using the classical approach with the true and mis-specified model, respectively. \texttt{MED\_Wt\_Orig} and \texttt{MED\_Wt\_Mis} denote the MED estimated using the RR approach with the true and mis-specified model, respectively.}
 \label{fig:RRsigEmaxEmax}
\end{figure}

For the first set of simulations (data generated from Table~\ref{tab:SimulationTable}), the results are shown for $n = 25$. Figure \ref{fig:RR} and Figure \ref{fig:IRNLS} give the boxplots of estimated MED and the summary statistics of the relative deviation, where the RR and IRWLS approach are respectively applied with the true dose-response shape. Figure \ref{fig:Emax} and Figure \ref{fig:Emax2} show the corresponding distribution of the MED estimates for the emax model in Table \ref{tab:SimulationTable} and Figure \ref{fig:sigEmax} and Figure \ref{fig:sigEmax2} for the sigmoidal Emax model. From Figure \ref{fig:RR}, it is evident that the MED estimates, under the true model fit with the new approaches, do not deviate much from the classical approach. However, some of the weights performed better compared to the others in obtaining the MED estimates. From Figure \ref{fig:Emax} it is observed that the weights $w_1$, $w_4$, $w_5$ and $w_6$ are giving similar or better estimates than the classical approach when the emax model is fitted with the RR approach. From Figure \ref{fig:sigEmax} one can conclude that the weights $w_3$, $w_4$, $w_5$ and $w_6$ are giving similar estimates as the classical approach when the sigEmax model is fitted with the RR approach. For the IRWLS approach, the MED estimates have more or less identical distribution as the classical approach for the sigmoidal Emax model (see Figure \ref{fig:sigEmax2}). However, in Figure \ref{fig:Emax2}, we observe that the IRWLS approach leads to biased estimates of MED for the weights: $w_1$, $w_2$, $w_5$ and $w_6$. Hence, in the evaluation of the performance of the different approaches under model misspecification we only applied the RR approach. The results are shown in Figure \ref{fig:RRsigEmaxLin} and \ref{fig:RRsigEmaxEmax}. Additionally, only the weights $w_4$, $w_5$ and $w_6$ are considered under model mis-specification because these weights show consistent behavior under the true model fit as can be seen from Figure \ref{fig:Emax} and \ref{fig:sigEmax}.

Figure \ref{fig:RRsigEmaxLin} shows the distribution of the MED estimates when the RR approach with weight $w_6$ is applied with a linear model to the data simulated from the sigmoidal Emax model in Table \ref{tab:SimulationTable}. Similarly, Figure \ref{fig:RRsigEmaxEmax} shows the distribution of the MED estimates when the RR approach with weight $w_5$ is applied with an emax model to the data simulated from the sigmoidal Emax model in Table \ref{tab:SimulationTable}. It is interesting to note from Figure \ref{fig:RRsigEmaxLin} that the weighted linear fit estimates the MED as accurate as the MED estimated under the true model (sigmoidal Emax). Furthermore, as the sample size increases from $25$ patients to $50$ patients per dose group, the precision in estimating the correct MED with the weighted linear fit increases. This is evident from 
Figure \ref{fig:sigEmaxlinear1} and \ref{fig:sigEmaxlinear2}. 
Moreover, with the true model fit (sigmoidal Emax), the performance of the RR approach is similar to the classical approach (see Figure \ref{fig:RRsigEmaxLin}). The weighted linear model was also fitted with $w_4$ and $w_5$. But with $w_4$, the RR approach is performing identical to the classical approach and with $w_5$, it is giving similar performance as $w_6$. Hence, the respective plots corresponding to weights $w_4$ and $w_5$ are added to Appendix~\ref{Appendix A.3}. From Figure \ref{fig:RRsigEmaxEmax} we observe that the RR approach can improve the bias in estimating the true MED over the classical approach under both the misspecified (emax) and true (sigmoidal Emax) models. For the latter case, this is contradictory to the presumed theoretical assumptions that the classical approach, which is based on the maximum likelihood-based estimates, already provides the UMVUE estimates for the model parameters. Hence, it is non-intuitive that the RR approach achieves any improvement over the classical approach under the true model assumptions. However, our observations can be explained by the fact that for non-linear models the predictions can be mean- and median-biased even when the model family has been correctly specified and that a weighted least squares fit has still the potential to reduce this bias for the doses of interest (i.e. those close to the target dose).  The weighted emax model is also fitted with weights $w_4$ and $w_6$ and provided similar results as $w_5$. The corresponding plots are added to the Appendix~\ref{Appendix A.4}.  We can infer from above that for mis-specified models, one can reduce the bias considerably and may inflate the variance negligibly. For the true model fit, the RR approach either performs similar to the classical approach or achieves a small reduction in bias at the cost of marginal inflation of the variance. In summary, there is practically no harm in applying the new method for MED estimation with an appropriate choice of the weight function.

So far, we have observed that the RR approach succeeds in estimating the MED more accurately than the existing approaches for data generated from two popular data generating shapes. With the next set of simulations, we intend to test if we can enhance the dose-ﬁnding step of the existing dose-response analysis strategies by integrating them with the RR approach.  

\begin{figure}%
 \centering
\subfigure[Evaluating MED estimation accuracy between the classical MCP-Mod approach and the RR approach combined with MCP-Mod approach for data simulated from the linear model]{%
\includegraphics[scale=0.76]{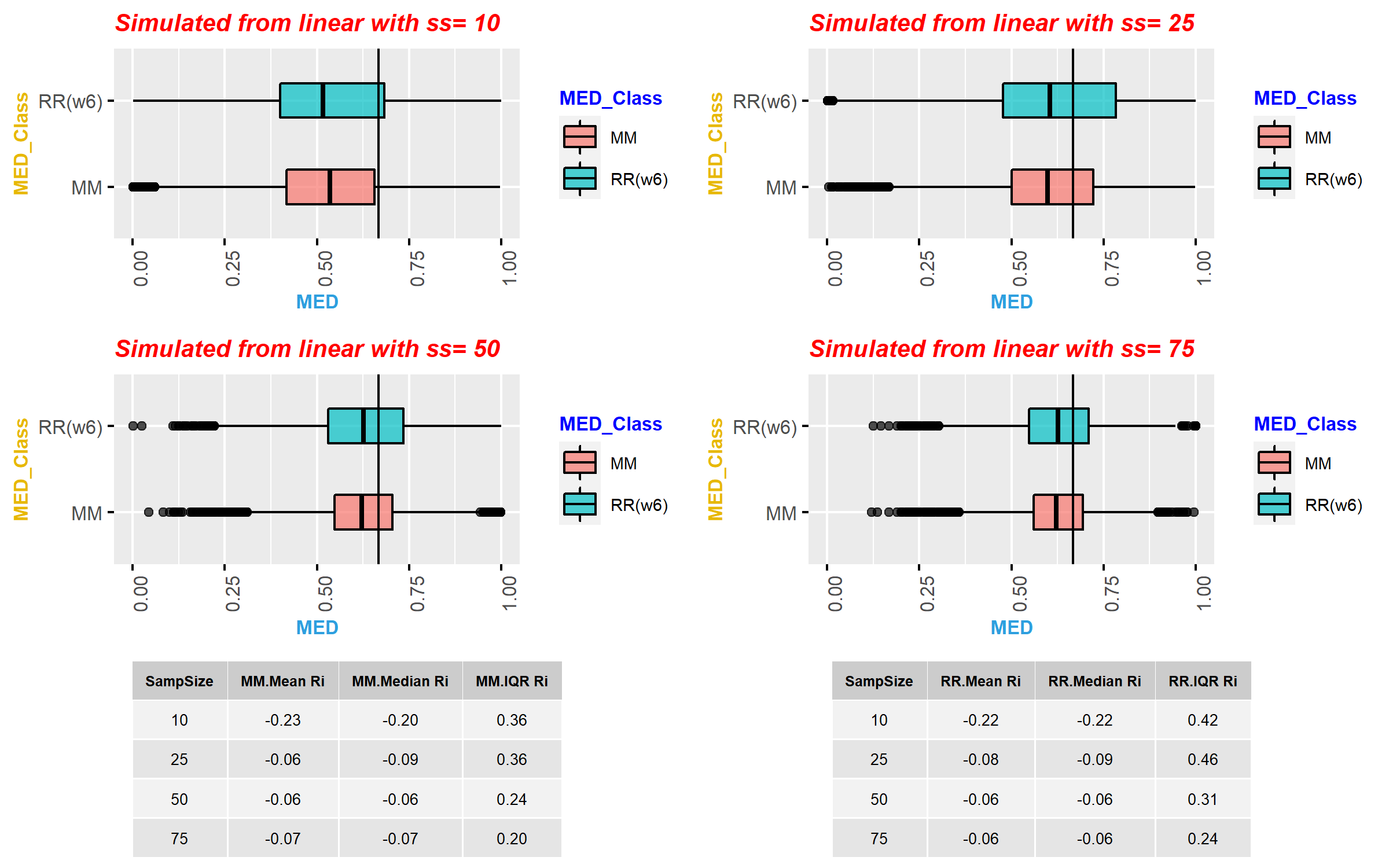}
\label{fig:RR-MM1a}}
\qquad
\subfigure[Evaluating MED estimation accuracy between the classical MCP-Mod approach and the RR approach combined with MCP-Mod approach for data simulated from the linear log model]{%
\includegraphics[scale=0.76]{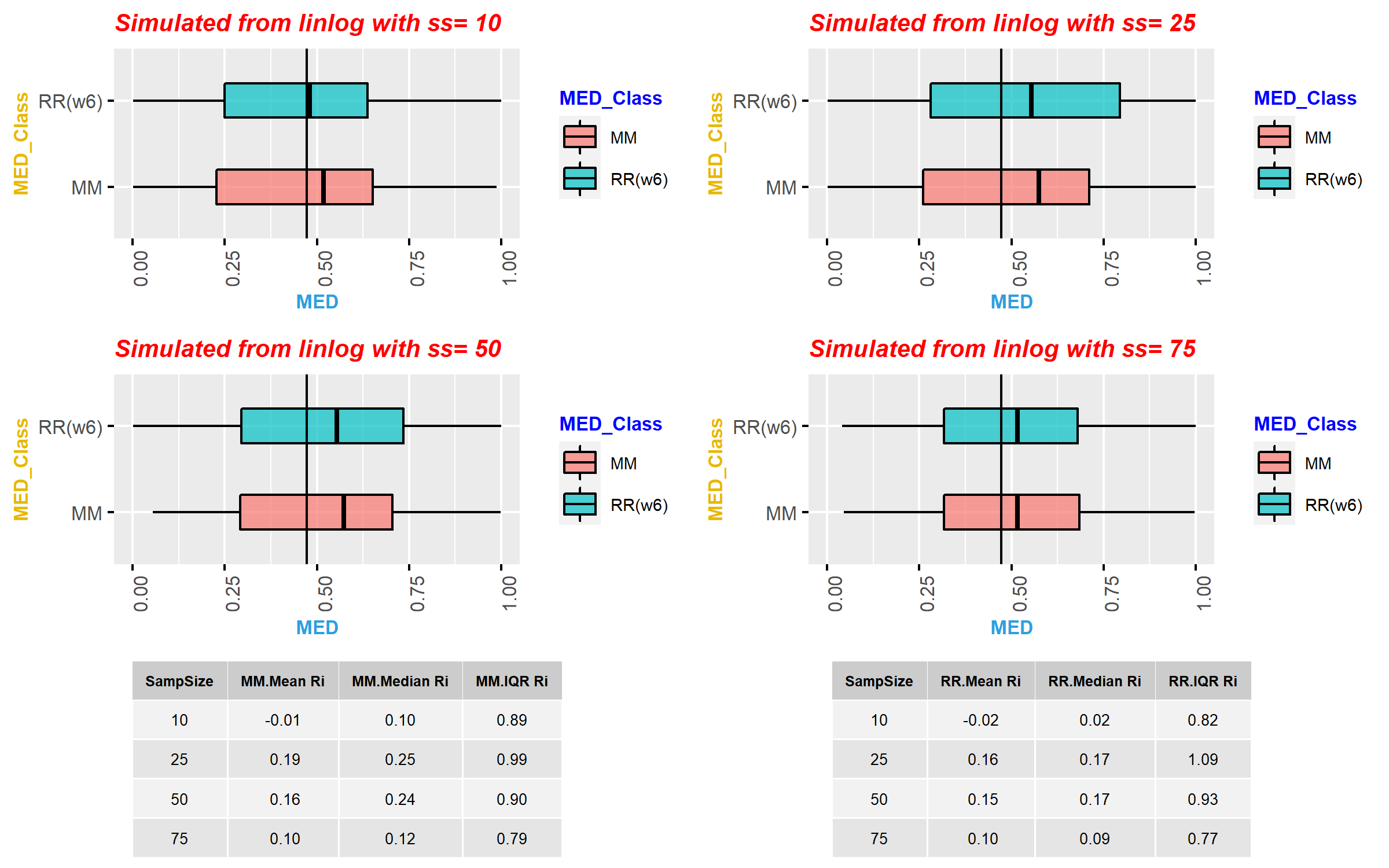}
\label{fig:RR-MM1b}}
\caption{Boxplot distribution of MED estimated under the classical MCP-Mod and RR combined with the MCP-Mod approach with weight $w_6$. This is followed by a table which summarizes the mean, median and IQR of $R_i$ for the $2000$ simulation trials. Figure (a) and Figure (b) show the comparison of MED estimation accuracy between the two approaches for data simulated from the linear and linear log model in Table~\ref{tab:SimulationTableMM-RR}, respectively, with sample size $10$, $25$, $75$ and $50$. \texttt{RR(w6)} and \texttt{MM} denote the MED estimated using the RR integrated with MCP-Mod approach and the classical MCP-Mod approach respectively. }
 \label{fig:RR-MM1}
 
\end{figure}

\begin{figure}%
 \centering
 \subfigure[Evaluating MED estimation accuracy between the classical MCP-Mod approach and the RR approach combined with MCP-Mod approach for data simulated from the emax model]{%
\includegraphics[scale=0.76]{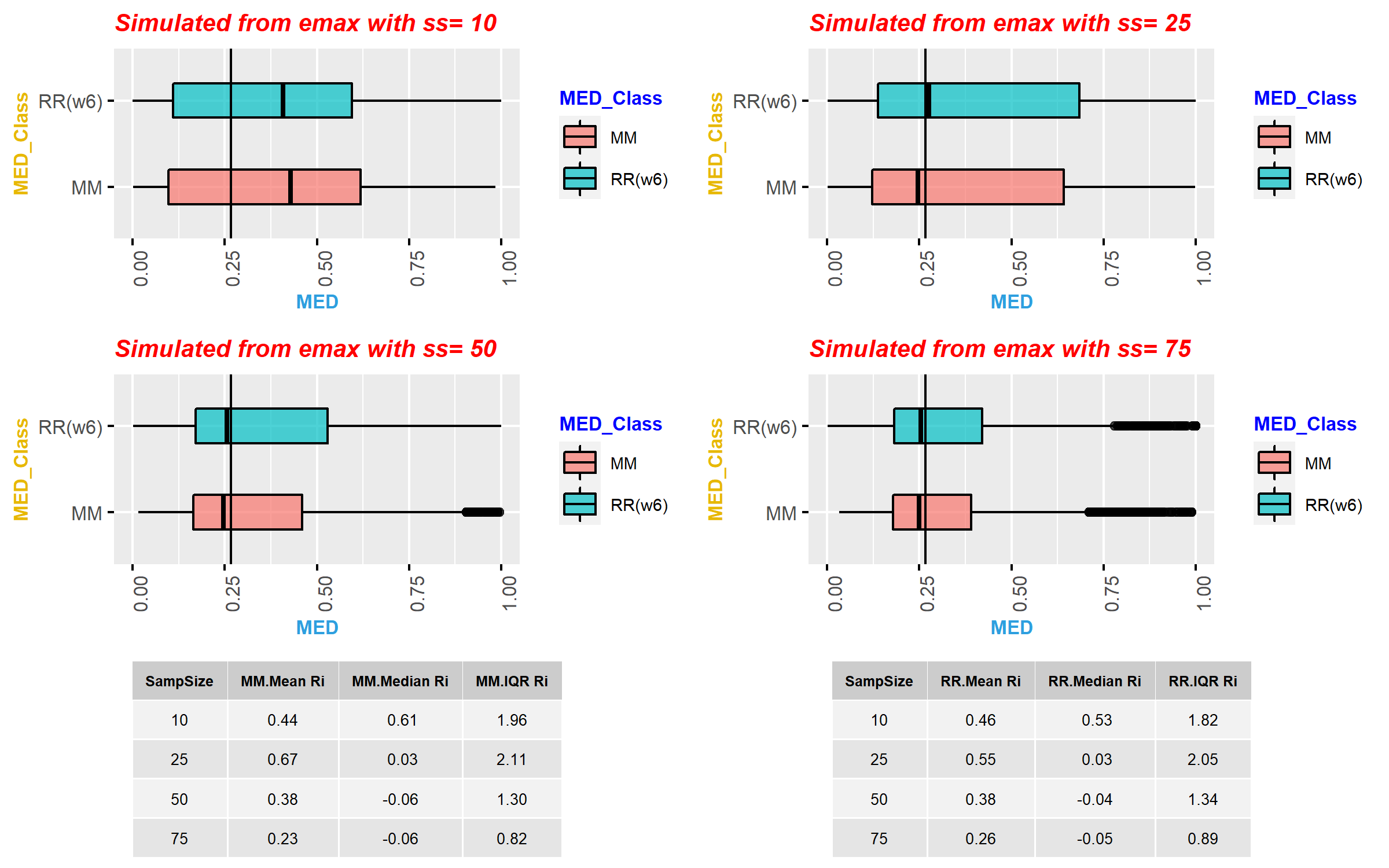}
 \label{fig:RR-MM2a}}
\qquad
\subfigure[Evaluating MED estimation accuracy between the classical MCP-Mod approach and the RR approach combined with MCP-Mod approach for data simulated from the emax2 (second emax) model]{%
\includegraphics[scale=0.76]{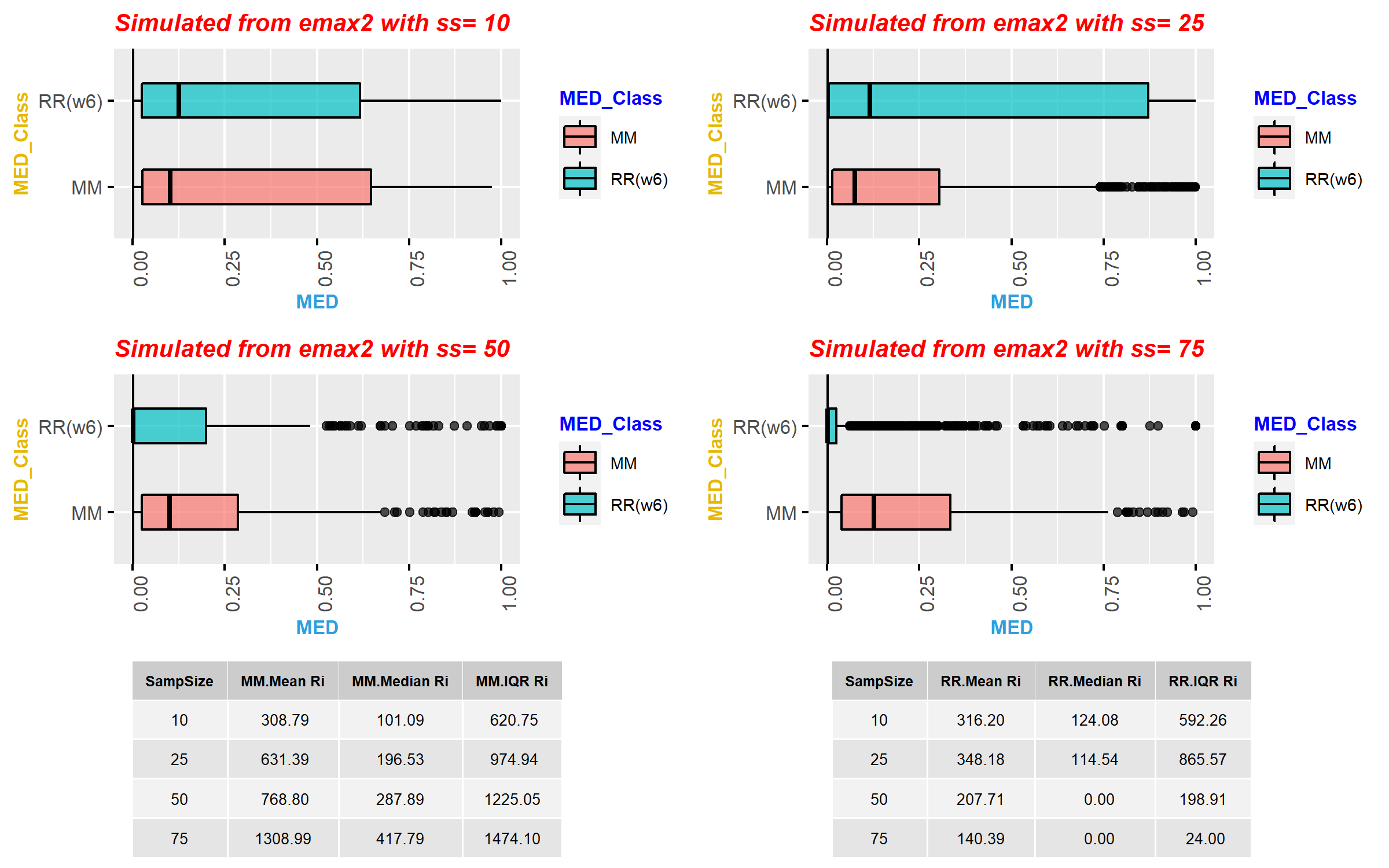}
 \label{fig:RR-MM2b}}
\caption{Boxplot distribution of MED estimated under the classical MCP-Mod and RR combined with the MCP-Mod approach with weight $w_6$. This is followed by a table which summarizes the mean, median and IQR of $R_i$ for the 2000 simulation trials. Figure (a) and Figure (b) show the comparison of MED estimation accuracy between the two approaches for data simulated from the  emax and emax2 in Table~\ref{tab:SimulationTableMM-RR}, respectively, with sample size $10$, $25$, $75$ and $50$. \texttt{RR(w6)} and  \texttt{MM} denote the MED estimated using the RR integrated with MCP-Mod approach and the classical MCP-Mod approach respectively. }
 \label{fig:RR-MM2}
\end{figure}

\begin{figure}%
 \centering
 \subfigure[Evaluating MED estimation accuracy between the classical MCP-Mod approach and the RR approach combined with MCP-Mod approach for data simulated from the sigmoidal Emax model]{%
\includegraphics[scale=0.76]{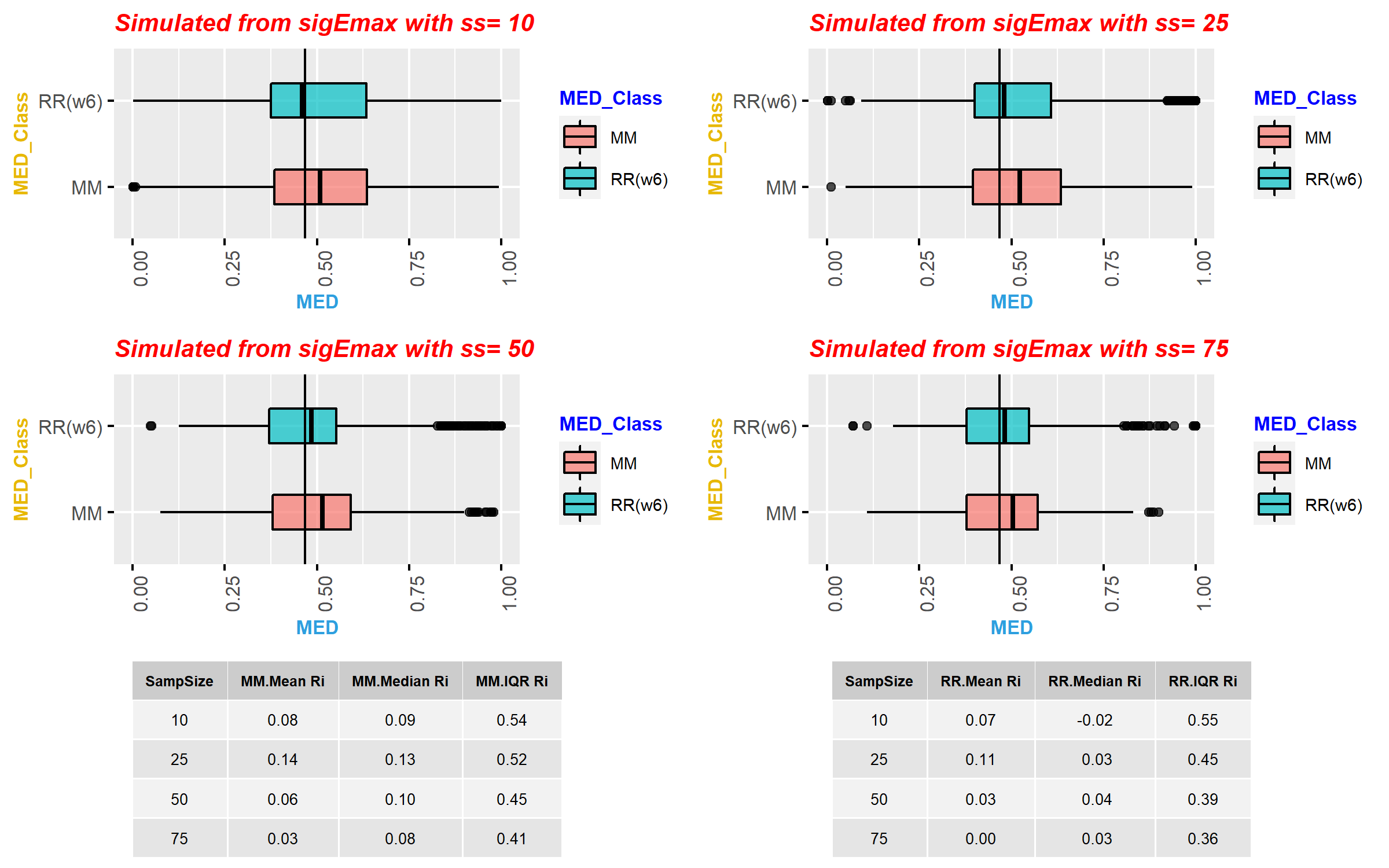}
 \label{fig:RR-MM3a}}
\qquad
\subfigure[Evaluating MED estimation accuracy between the classical MCP-Mod approach and the RR approach combined with MCP-Mod approach for data simulated from the power model]{%
\includegraphics[scale=0.76]{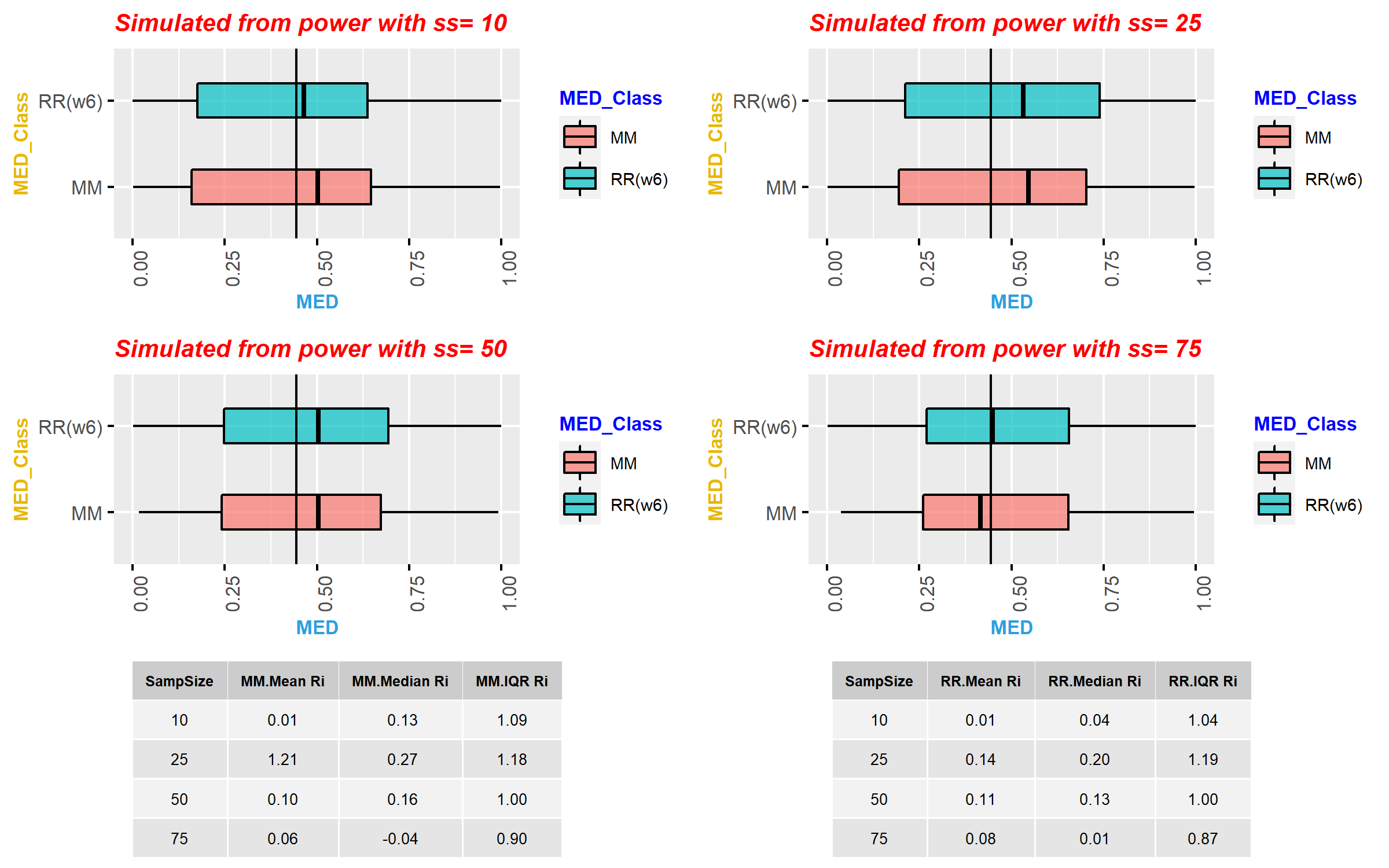}
\label{fig:RR-MM3b}}

\caption{Boxplot distribution of MED estimated under the classical MCP-Mod and RR combined with the MCP-Mod approach with weight $w_6$. This is followed by a table which summarizes the mean, median and IQR of $R_i$ for the $2000$ simulation trials. Figure (a) and Figure (b) show the comparison of MED estimation accuracy between the two approaches for data simulated from the sigmoidal Emax and power model in Table~\ref{tab:SimulationTableMM-RR}, respectively, with sample size $10$, $25$, $75$ and $50$. \texttt{RR(w6)} and \texttt{MM} denote the MED estimated using the RR integrated with MCP-Mod approach and the classical MCP-Mod approach, respectively. }
 \label{fig:RR-MM3}
\end{figure}

\begin{figure}%
 \centering
\includegraphics[width=\textwidth]{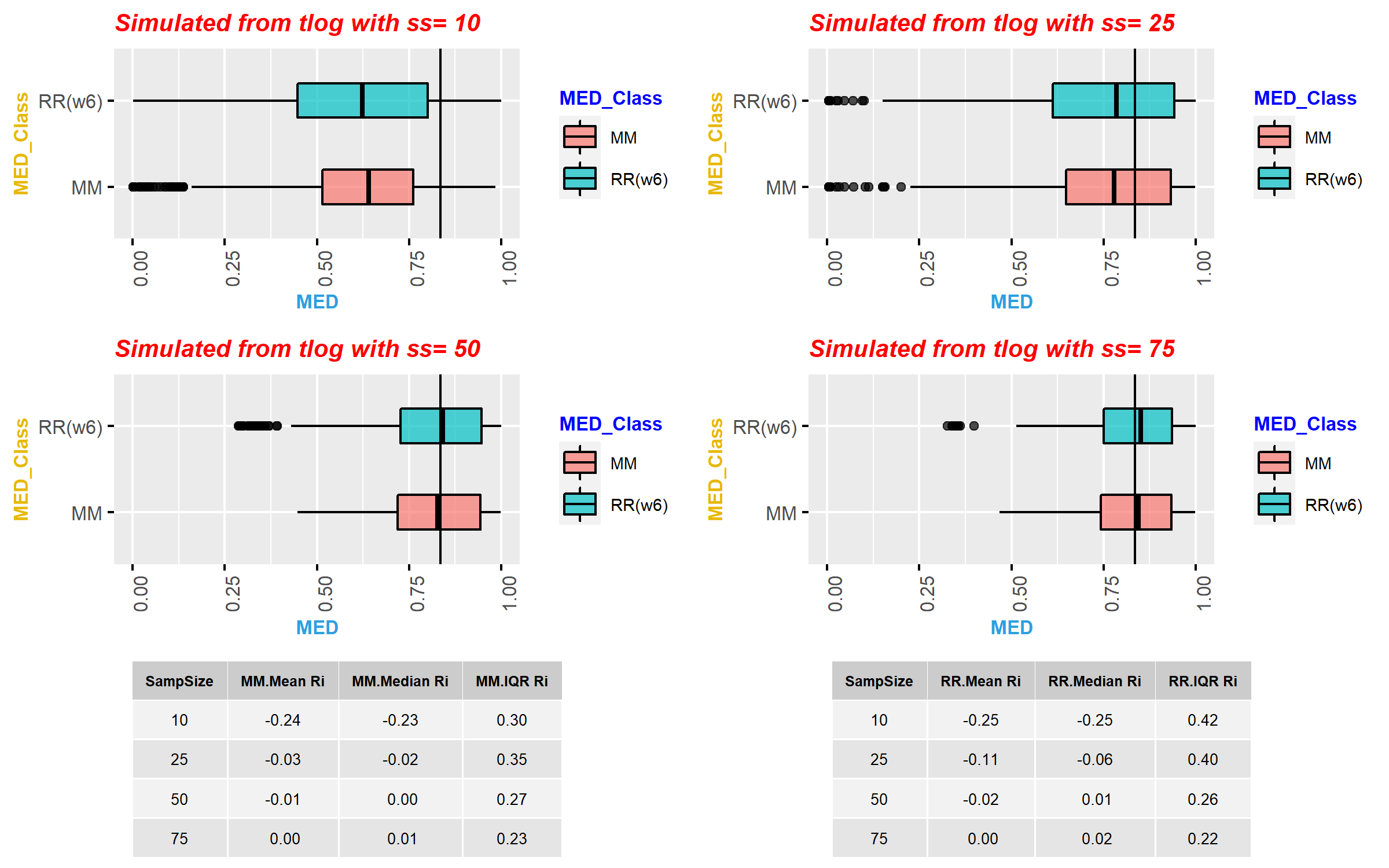}
\caption{Boxplot distribution of MED estimated under the classical MCP-Mod and RR combined with the MCP-Mod approach with weight $w_6$ for data simulated from the truncated logistic model. This is followed by a table which summarizes the mean, median and IQR of $R_i$ for the $2000$ simulation trials. The figure compares the accuracy in estimating the true MED between the two approaches data simulated from the truncated logistic model in Table~\ref{tab:SimulationTableMM-RR} with sample size $10$, $25$, $75$ and $50$. \texttt{RR(w6)} and \texttt{MM} denote the MED estimated using the RR integrated with MCP-Mod approach and the classical MCP-Mod approach respectively. }
 \label{fig:RR-MM4}
\end{figure}

As stated in Section~\ref{Simulation}, for simulation scenario 2, data are simulated from seven non-linear models shown in Table~\ref{tab:SimulationTableMM-RR} and  Table~\ref{tab:SimulationTable_newMM-RR} to evaluate the effectiveness of the RR approach when combined with the MCPMod approach.  In this simulation scenario, the linear, emax and the sigEmax model highlighted in blue in Table~\ref{tab:SimulationTableMM-RR} is considered in the candidate set of MCPMod. Since the simulation scenarios in the two tables are leading to similar results, we are showing the results only for the scenarios from Table~\ref{tab:SimulationTable_newMM-RR} in this section. Note that the second emax model, emax2, is not desirable as it does not reach the anticipated efficacy. It is introduced in Saha and Brannath, \cite{saha2018comparison} and here, to evaluate the performance of MCPMod under unfavourable situations. Data are also simulated from the power and truncated logistic  (tlog) models apart from the models in the candidate set because we want to check how accurately the MCPMod-RR approach estimates the true MED under model misspecification. Figure~\ref{fig:RR-MM1} to Figure~\ref{fig:RR-MM4} shows the performance of the aforementioned approaches for data simulated from the models in Table~\ref{tab:SimulationTable_newMM-RR}. As evident from Figure~\ref{fig:RR-MM1}, we do not gain much with the MCPMod-RR approach for data simulated from the linear model. We get the same accuracy as the classical MCPMod approach but the variance is slightly inflated under the MCPMod-RR approach. However, amongst models included in the candidate set, we see a considerable improvement in the MED estimation for the sigmoidal Emax model with the MCPMod-RR approach, as evident from Figure ~\ref{fig:RR-MM3a}. The improvement is significant both in terms of accuracy and precision. For data generated from emax model, the MCPMod-RR approach is showing comparable performance as the original MCP-Mod approach (see Figure~\ref{fig:RR-MM2a}). Particularly interesting is the performance for the second emax model (emax2) in Figure~\ref{fig:RR-MM2b}. It is important to mention here that in Saha and Brannath, \cite{saha2018comparison} MCPMod fails to detect the emax2 dose-response shape well and so the MED estimation with MCPMod was also very poor for the data generated from the emax2 model. Note that the second emax model (emax2) is quite close to a constant model with no dose-response effect and the MED occurs at the placebo dose, $d=0$. We observe in Figure~\ref{fig:RR-MM2b} that though the MCPMod-RR approach is not able to accurately estimate the true MED for small sample sizes, but as the sample size increases the accuracy and precision in estimating the true MED drastically improves compared to the classical MCPMod approach. For the models not included in the candidate set; while the accuracy in estimating the true MED improves for the linear log model (see Figure~\ref{fig:RR-MM1b}) and the power model (see Figure~\ref{fig:RR-MM3b}) with the MCPMod-RR approach, the performance of the classical MCP-Mod approach and the MCPMod-RR approach remains comparable for the truncated logistic model (see Figure~\ref{fig:RR-MM4}). It is important to point out here that for large sample sizes (sample size $= 50$ and $75$ per dose group), the MCPMod-RR approach is showing improvement or similar performance as the classical MCP-Mod approach both in terms of accuracy and precision across all the shapes. For small sample sizes, this is not always true. Often the MCPMod-RR method estimates the true MED at the cost of inflating variance for small sample sizes. 

After analyzing the estimation performance of the new approaches, the next goal is to analyze the coverage probability of the confidence intervals for the MED. In order to evaluate the coverage probability of the different methods, we conduct simulation studies whose set-up was already discussed earlier in Section \ref{Simulation}. Table \ref{tab:coverage_Emax} gives the coverage probabilities of the different
methods for the Emax model, and Table \ref{tab:coverage_sigEmax1} and \ref{tab:coverage_sigEmax2} give the corresponding results for the sigmoidal Emax model. Note that the RR and IRWLS methods are applied with weights $w_5$ and $w_6$ for evaluating the coverage probability. However, with the RR approach, we were unable to derive the confidence bound for the MED with weight $w_6$ for the sigEmax data generating shape in $10\%$ simulations. This is mainly because in many situations the covariance matrix is not estimable due to singularity issues. Hence, we show results for the RR implementation only with $w_5$ in Table \ref{tab:coverage_sigEmax1} and \ref{tab:coverage_sigEmax2}. For IRWLS, results are only shown for weight $w_6$ for both the emax and sigmoidal Emax model because $w_5$ is giving similar performance as $w_6$. From the results in Table \ref{tab:coverage_Emax}, it is clear that the profile likelihood and percentile bootstrap approach perform superior to all the other methods for the Emax model. For sample sizes larger than $25$, the percentile bootstrap performs even better than the profile likelihood approach. The Weighted RR approach like the classical approach fails to attain the nominal value. The IRWLS method does not perform well here. They lead to undercoverage in most of the situations under the emax model. The poor performance of the IRWLS approach may be due to the fact that theoretically the asymptotic covariance of the weighted least squares estimates is given by the covariance matrix in equation (\ref{Var_MED_RR}) and the approximate asymptotic covariance given by the IRWLS approach (equation (\ref{ConfIntIRNLS})) in Section \ref{IRWLS} deviates from the covariance in equation (\ref{Var_MED_RR}). The results in Table \ref{tab:coverage_sigEmax1} show that the percentile bootstrap approach is attaining the nominal level but the classical approach and RR approaches show overcoverage for the sigmoidal Emax model for large sample sizes. This may be due to the flexibility of the sigmoid Emax curve when a limited number of doses are included in the study design. For a limited number of dose groups, the study data cannot discriminate very well between different possible model fits and thereafter end up in very wide confidence intervals in most of the cases. The IRWLS and profile likelihood approaches are prone to undercoverage for all sample sizes. However, the coverage of the confidence interval also depends on the dose allocation design. As discussed in Baayen and Hougaard, \cite{baayen2015confidence} and Dette et al., \cite{dette2008optimal} the coverage of the confidence bounds is expected to improve if a more optimal dose range is investigated (optimal in terms of dose allocation). To investigate this, we ran few more simulations for the sigmoidal Emax model in Table \ref{tab:SimulationTableEC} with dose levels
$0,\, 0.25,\, 0.5,\, 0.75,\, \text{and } 1$. The coverage probabilities based on $2000$ simulated datasets are shown in Table \ref{tab:coverage_sigEmax2}. The results clearly improve. The classical approach and the weighted RR approach both attain the nominal level with the new dose allocation design. The percentile bootstrap approach is attaining the nominal level similar to the earlier scenario. But the profile likelihood method is prone to undercoverage in this simulation scenario as well. The undercoverage of the profile likelihood confidence interval might be because of the convergence issues arising in the implementation of the method. We need to investigate more to understand this properly. The performance of IRWLS method improves but it fails to attain the nominal level even in this simulation scenario. The simulations in Table \ref{tab:coverage_sigEmax2} consider an ideal case where the dose-response relationship is quite clear, and an informative dose range is included in the study. 

We also investigate the coverage probabilities of the different approaches under model mis-specification. The data are simulated from the sigmoidal Emax model in Table \ref{tab:SimulationTableEC} and fitted using an unweighted and weighted linear model (with weight ($w_6$) as in Figure \ref{fig:RRsigEmaxLin}) model and an unweighted and weighted emax model (with weight $w_5$ as in Figure \ref{fig:RRsigEmaxEmax}). For the unweighted linear and emax model, confidence intervals are estimated using the classical approach, bootstrap approach, and profile likelihood approach. For the weighted linear and emax model, confidence intervals are estimated using the RR approach. Coverage probabilities of the $95\%$ confidence interval from the linear and emax fit are shown in Table  \ref{tab:coverage_sigEmaxlinear1} and \ref{tab:coverage_sigEmaxEmax}, respectively. We observe in Table \ref{tab:coverage_sigEmaxlinear1} that the classical approach tend to give overcoverage, whereas the percentile bootstrap and the profile likelihood approach tends to give undercoverage. The RR approach attains the nominal level of $95\%$ with the weighted linear fit. Similarly, Table \ref{tab:coverage_sigEmaxEmax} shows that the RR approach with the weighted emax fit is performing better than the classical approach, percentile bootstrap approach and profile likelihood approach with the unweighted emax fit. However, for large sample sizes even the RR approach is giving slight overcoverage as can be seen from Table  \ref{tab:coverage_sigEmaxlinear1} and \ref{tab:coverage_sigEmaxEmax}.

\begin{table}[htb!]
\centering
\caption{Table showing the coverage of $95\%$ confidence interval under the different methods and across different sample sizes per dose group for data simulated from the emax model in Table \ref{tab:SimulationTableEC}}
\label{tab:coverage_Emax} 
\begin{tabular}{c|cccccc}
     \toprule
    \multirow{2}{*}{Sample Size} & \multicolumn{6}{c}{Methods}\\
    \cmidrule{2-7}
    & Classical &  RR($w_6$)&  RR($w_5$)& IRWLS ($w_6$)& PBootstrap & Prof Lik\\
      \midrule
     25&0.8292&0.8466&0.8390&0.6720& 0.9355 &0.9375  \\
     50&0.8784&0.9207&0.9035&0.7420& 0.9500 & 0.9415 \\
     75&0.8988&0.9265&0.9235&0.7510& 0.9580& 0.9525 \\
     100&0.9095&0.9378&0.9255&0.7675&0.9550 &0.9545 \\
      \hline
     \end{tabular}
\end{table}

\begin{table}[htb!]
\centering
\caption{Table showing the coverage of $95\%$ confidence interval under the different methods and across different sample sizes per dose group for data simulated from the sigEmax model in Table \ref{tab:SimulationTableEC}}
\label{tab:coverage_sigEmax1} 
\begin{tabular}{c|ccccc}
     \toprule
    \multirow{2}{*}{Sample Size} & \multicolumn{5}{c}{Methods}\\
    \cmidrule{2-6}   
    & Classical & RR($w_5$)& IRWLS($w_5$)& PBootstrap & Prof Lik\\
      \midrule
     25&  0.9545  &0.9070 & 0.6845  & 0.9550  & 0.9132 \\
     50& 0.9735& 0.9480  &  0.6375 & 0.9525 & 0.8993 \\
     75&0.9835&  0.9715  & 0.6890 & 0.9520 & 0.9049\\
     100&0.9865& 0.9785  & 0.6570  &0.9510 & 0.9118 \\
      \hline
     \end{tabular}
\end{table}

\begin{table}[htb!]
\centering
\caption{Table showing the coverage of $95\%$ confidence interval under the different methods and across different sample sizes per dose group for data simulated from the sigEmax model with dose groups $0, 0.25, 0.5, 0.75$ and $1$ in Table \ref{tab:SimulationTableEC}}
\label{tab:coverage_sigEmax2} 
\begin{tabular}{c|ccccc}
     \toprule
    \multirow{2}{*}{Sample Size} & \multicolumn{5}{c}{Methods}\\
    \cmidrule{2-6}
    & Classical &  RR($w_5$)&  IRWLS($w_5$)& PBootstrap& Prof Lik\\  
      \midrule
     25& 0.9425   &0.9290 & 0.8015   &0.9370& 0.9097  \\
     50&0.9595& 0.9485  &  0.8315  &0.9500& 0.9410 \\
     75&0.9550 & 0.9500  & 0.8435 &  0.9500&0.9317   \\
     100&0.9580& 0.9500 & 0.8565  &0.9475& 0.9276  \\
      \hline
     \end{tabular}
\end{table}
\begin{table}[htb!]
\centering
\caption{Table showing the coverage of $95\%$ confidence interval under the different methods and across two sample sizes per dose group for data simulated from the sigEmax model in Table \ref{tab:SimulationTableEC} and fitted on linear model}
\label{tab:coverage_sigEmaxlinear1} 
\begin{tabular}{c|cccc}
     \toprule
    \multirow{2}{*}{Sample Size} & \multicolumn{4}{c}{Methods}\\
    \cmidrule{2-5}
    & Classical & RR($w_6$)& PBootstrap& Prof Lik\\
      \midrule
     25& 0.9800   &  0.9530&0.903& 0.9265  \\
     50&0.9865 &0.9665&0.891& 0.9180 \\
       \hline
     \end{tabular}
\end{table}
\begin{table}[htb!]
\centering
\caption{Table showing the coverage of $95\%$ confidence interval under the different methods and across two sample sizes per dose group for data simulated from the sigEmax model in Table \ref{tab:SimulationTableEC} and fitted on emax model}
\label{tab:coverage_sigEmaxEmax} 
\begin{tabular}{c|cccc}
     \toprule
    \multirow{2}{*}{Sample Size} & \multicolumn{4}{c}{Methods}\\
    \cmidrule{2-5}
    & Classical & RR($w_5$)& PBootstrap& Prof Lik\\
      \midrule
     25& 0.9235   &  0.9470&0.8920& 0.9705   \\
     50&0.9470 &0.9690&0.8915&0.9720 \\
       \hline
     \end{tabular}
\end{table}

The percentile bootstrap approach performs superior to the other approaches for all the scenarios where the true underlying model is fitted. Unlike other approaches, it is not so sensitive to the dose-allocation and attains the nominal level in all the scenarios. Amongst the new approaches suggested in this article, the RR approach performs better than the IRWLS approach. It performs similar to the classical approach when the true model is fitted but it performs better than the other approaches under model mis-specification. One disadvantage of the percentile bootstrap approach and profile likelihood approach is that they are grid-based algorithms and the inference from these approaches are heavily dependent on the choice of grids. Furthermore, due to the repeated optimization involved in these grid-based methods, they are quite time-consuming as compared to the other methods. The advantage of the RR approach is that it is comparatively much less time-consuming.

\section{Discussion}
\label{Discussion}
While developing a drug, typically the choice of dose is based on a Phase II dose‐finding trial, where selected doses are included along with the placebo. The analysis of such dose-ﬁnding trial/studies can be classiﬁed into two approaches: separate comparisons of each dose to placebo using the multiple comparison procedure (MCP) or a model‐based strategy, where a dose-response model is fitted to the data \cite{pinheiro2006analysis,bates1988nonlinear}. The first approach is preferred when there are only a few dose levels in a dose-ﬁnding study and typically limited knowledge is available about the dose-response relationship. But if there exist several dose groups with multiple observations and we want to infer beyond the observed doses, modeling techniques should be adopted. In this article we have focused on target dose estimation, particularly, MED, using model-based strategies. Our inspiration behind the method proposed in this article lies in our earlier article, \cite{saha2018comparison}. While comparing the dose-response profile estimation strategies in the above article, we inferred that while the existing approaches can very well establish the PoC of no dose-related effect and approximately estimate the true dose-response shape by a sufficiently well-approximated model, there is still some scope of improvement in the target dose estimation approach. 

Note that in the process of obtaining the approximate dose-response shape in the existing dose-response profile estimation methods, parameter estimates are obtained that focus on accurately estimating the entire dose-response shape. The classical approach uses these parameter estimates and it fails to provide accurate target dose estimates when the true underlying model is misclassified \cite{saha2018comparison}. This is mainly because, in the course of estimating the entire dose-response curve accurately, it loses in precision and gains bias in estimating the dose-response curve in a particular region around the target dose. In this article, inspired by the above notion, two novel approaches (IRWLS and RR) are introduced for estimating the MED and its confidence interval. Both approaches aim to provide solutions to a weighted least squares minimization, where the doses around the MED are given more weights than the doses away from the MED. In the first approach, we apply an iterated re-weighted least squares algorithm to obtain our estimates and in the second approach, we propose to solve a weighted sum of scores function. With the latter approach, we obtain consistent and asymptotic normal estimates under specific regularity conditions, when the model family is correctly specified. Since the RR approach performs much better in our simulations than the IRWLS approach, we propose to use the second and discard the IRWLS approach, which often fails to reach the target coverage probability. We also observe from our simulations that the RR approach gives results similar to the classical approach when the true model is fitted. Under model misspecifications, the RR approach produces less biased estimates of the MED compared to the classical approach. A limitation of the RR approach is that the asymptotic theory relies on the assumption of a correctly specified model and needs to be extended under model misspecifications which is a difficult task, as, consistency cannot be expected in general. We have conducted experiments where the RR approach is being applied for MED estimation after the model selection by MCPMod approach. We have shown with our simulations that the RR approach indeed enhances the dose-finding step of the MCPMod approach and it might be advisable to use it for target dose estimation for future stages of drug development.

Though the presentation of confidence intervals is quite mandatory in the area of dose-finding, the methodology for this has not been well explored in the literature. Helms et al., \cite{helms2015point} studied design with active control of a new drug with several dose levels and investigated the estimation of the target dose that led to the same efficacy as the active control. The authors suggested ways to construct confidence intervals under the assumption of a linear dose-response curve and normal errors. Dilleen et al., \cite{dilleen2003non} investigated a similar design with a similar objective as Helms et al., \cite{helms2015point} but they also considered a placebo control group in their study design. They assumed monotonically increasing dose-response functions and presented different non-parametric methods to estimate the monotonic dose-response curve. They also suggested ways to construct confidence intervals for the dose-response curve using the isotonic approach from Korn, \cite{korn1982confidence} and bootstrap approaches. But both these approaches are using a design different from ours. We do not have an active control in our design and our objective is to estimate a dose that produces the desired threshold (which is known apriori) over the placebo. Traditionally the classical approach or MCP approaches were used to address this issue. The approaches for confidence bounds by Baayen and Hougaard \cite{baayen2015confidence} are something new in this area and they introduced a wide spectrum of options by which one can estimate the confidence bounds for the target dose of interest. We have only discussed the percentile bootstrap approach and profile likelihood approach from their article but they have further implemented two other bootstrap approaches. Unfortunately, they were not performing well, so we have not included them in our simulations. To the best of our knowledge, we did not find any article discussing and comparing the coverage of MED using modeling techniques.
\section*{Acknowledgements}

This work was supported by funding from the European Union’s Horizon 2020 research and innovation program IDEAS ("Improving Design, Evaluation and Analysis of early drug development Studies"; \url{www.ideas-itn.eu}) under the Marie Sklodowska-Curie grant agreement No 633567. 

\newpage
\section{Appendix}
\label{Appendix} 


\subsection{Justification behind the choice of loss function in the Robust Regression Approach}
\label{Appendix A.1}
In the weighted regression approach introduced in this article the primary objective is to minimize the following weighted SSE:

  \begin{equation*}
   \label{WeightedSSE}
       wSSE=\sum_{i,j}w_{ij}(d_i,MED(\boldsymbol{\theta}))(Y_{ij}-\mu(d_i,\boldsymbol{\theta}))^2
   \end{equation*}
   
   Ideally such a least square problem is dealt using an iterated re-weighted least squares algorithm (IRWLS). It starts with some good initial values \cite{ruckstuhl2010introduction,cleveland1979robust}. Using the initial estimates, the weights are obtained and then with these weights $wSSE$ is minimized. and thereafter the subsequent estimates are obtained. Typically the IRWLS algorithm involves an iterative procedure where at each iteration the weights are derived using the estimates from the earlier iteration and then with these particular weights the $wSSE$ is minimized to obtain the next(or final) regression estimates. In detail, for the $n^{th}$ iteration step under the IRWLS algorithm one need to minimize the following:
   
   \begin{equation}
   \label{WeightedScore}
       \sum_{i,j}w_{ij}(d_i,MED(\boldsymbol{\hat{\theta}_{n-1}}))(Y_{ij}-\mu(d_i,\boldsymbol{\theta}_{n}))^2
   \end{equation}
   
 This is equivalent to solving the following normal equations:
\begin{equation}
   \label{WeightedSSE2}
       \sum_{i,j}w_{ij}(d_i,MED(\boldsymbol{\hat{\theta}_{n-1}}))\frac{\partial(Y_{ij}-\mu(d_i,\boldsymbol{\theta_{n}}))^2}{\partial \boldsymbol{\theta}}=0
   \end{equation}
The IRWLS approach leads to a sequence of $\{\boldsymbol{\hat{\theta}_{n}}\}$ which usually converges with respect to some criteria (already elaborated in section \ref{IRWLS}) by minimizing the above in the $n^{th}$ iteration. For the final convergent step, $\boldsymbol{\hat{\theta}_n}$ is very close to $\boldsymbol{\hat{\theta}_{n-1}}$ and obtaining solutions to (\ref{WeightedSSE2}) by IRWLS is fairly similar to solving for the normal equations:
    \begin{equation}
   \label{WeightedSSE3}
       \sum_{i,j}w_{ij}(d_i,MED(\boldsymbol{\theta}))\frac{\partial(Y_{ij}-\mu(d_i,\boldsymbol{\theta}))^2}{\partial \boldsymbol{\theta}}=0
   \end{equation}
 
 This justifies the choice of the loss function in (\ref{FraimanLoss}) of Section~\ref{RobustEstimation} for the robust regression approach.
 
 \subsection{Verifying the assumptions by \cite{fraiman1983general} in the Robust Regression Approach}
 \label{Appendix A.2}
 In the robust regression (RR) approach we have considered the following function:
\begin{equation}
\label{FraimanLoss}
    \phi(Y_{ij},d_{ij},\boldsymbol{\theta})=(Y_{ij}-\mu(d_{ij},\boldsymbol{\theta}))\cdot w_{d_{ij}}(d_{ij},\boldsymbol{\theta})\frac{\partial \mu(d_{ij},\boldsymbol{\theta})}{\partial\boldsymbol{\theta}}
\end{equation}
 
 We have substituted $d_{ij}$ by $d_i$ since $d_{ij}=d_i \forall \, j$. From our model set-up in section~\ref{Parametric set-up}, $Y_{ij} \sim \mathcal{N}(\mu(d_i,\boldsymbol{\theta_0}),\sigma^2)$. Hence, we can write $$\lambda_F(\boldsymbol{\theta_0})=E_F(\phi(Z,\boldsymbol{\theta_0}))=E_F\bigg{(}(Y_{ij}-\mu(d_i,\boldsymbol{\theta_0}))\cdot w_{ij}(d_i,\boldsymbol{\theta_0})\frac{\partial \mu(d_i,\boldsymbol{\theta_0})}{\partial \boldsymbol{\theta}}\bigg{)}=0$$ 
We need to show that $\phi$ defined above satisfies the assumptions $H1-H4$ and $N1$ mentioned in Section~\ref{RobustEstimation}. We verify the assumptions in the subsequent steps.
\begin{itemize}
\item[H1:]The weights can be so chosen (for instance the weights $w_1,\hdots,w_6$ shown in Section \ref{Weights}) such that $\phi$ is continuous. $\mu$ and $w_{ij}$ are continuously differentiable over the range of parameters used  in the model. So assumption $H1$ is satisfied.
\item[H2:] $F$ is continuous in  the experimental set-up (\ref{anovamodel}). $E_F(\phi(Y,D,\theta_0))=0$ is evident, where $D$ is the random dose variable, introduced in Section~\ref{Parametric set-up}. Need to show; $E_F(\frac{\partial \phi(Y,D,\theta_0)}{\partial \theta})$ is non singular.\\
$E_F\bigg{(}\frac{\partial \phi}{\partial \theta}\bigg{)}=
E_F\bigg{(}-\frac{\partial \mu(D,\theta_0)}{\partial \theta}\psi_2(D,\theta_0)\bigg{)} +E_F\bigg{(}\psi_1(Y-\mu(D,\theta_0))\frac{\partial \psi_2(D,\theta_0)}{\partial \theta}\bigg{)}$
            
where $\psi_1(Y-\mu(D,\theta_0))=(Y-\mu(D,\theta_0))$ and $\psi_2(D,\theta_0)=w(D,\theta)\bigg{(}\frac{\partial \mu(D,\theta)}{\partial \theta}\bigg{)}^\prime$. The second term is 0, when the model is correctly specified and close to 0 if it is approximated by a fairly similar model. So,

$E_F\bigg{(}\frac{\partial \phi}{\partial \theta}\bigg{)}=-\sum_{i,j} \frac{\partial g(d_i,\theta_0)}{\partial \theta} \cdot \psi_2(d_i,\theta_0)
=-\sum_{i,j}w_{i,j}(d_i,\theta_0)\frac{\partial g(d_i,\theta_0)}{\partial \theta}\cdot   \bigg{(}{\frac{\partial g(d_i,\theta_0)}{\partial \theta_0}}\bigg{)}^\prime$

Similar to non-linear least square method (\cite{sebernonlinear}) the product of the gradient function and its transpose is assumed to be non-singular here. The weights are also defined in Section \ref{Weights} such that they are always non zero for all dose groups. So from there it follows that $E_F\bigg{(}\frac{\partial\phi}{\partial \theta}\bigg{)}$ is non-singular in our set-up.
\item[H3:] The third assumption claims that there exist a compact ball of radius $r_0 \geq 0$ around the true parameter  $\boldsymbol{\theta_0}$ denoted by $B(\boldsymbol{\theta_0},r_0)$  s.t. $\underset{\boldsymbol{\theta} \in B(\boldsymbol{\theta_0},r_0)}{\sup }\|\phi(Y,d,\boldsymbol{\theta})\|$ and $ \underset{\boldsymbol{\theta} \in B(\boldsymbol{\theta_0},r_0)}{\sup } \|\frac{\partial (\phi(Y,d,\boldsymbol{\theta}))}{\partial \boldsymbol{\theta}}\|$ are both $F$- integrable.
  
It is sufficient to show; $\phi(Y,D,\boldsymbol{\theta})$ and $\frac{\partial (\phi(Y,D,\boldsymbol{\theta}))}{\partial \boldsymbol{\theta}}$ are bounded in the compact neighbourhood of $\boldsymbol{\theta}$.\\
  We have,
  $$\phi(Y_{i,j},d_i,\boldsymbol{\theta})=\psi_1(Y_{i,j},d_i,\boldsymbol{\theta})\psi_2(d_i,\boldsymbol{\theta})$$
 Since $\mu$ is continuously differentiable with respect to $\boldsymbol{\theta}$, its derivative is bounded in this compact neighbourhood. The weight functions are so chosen that they are also bounded by 1. So $\psi_2$ is bounded (say by $M$) and we have,
$$\mid\phi(Y_{i,j},d_i,\boldsymbol{\theta})\mid\leq M.\mid \psi_1(Y_{i,j},d_i,\boldsymbol{\theta})\mid$$
If $\exists \, r_0$ s.t. $\|\boldsymbol{\theta}-\boldsymbol{\theta_0}\|\leq r_0$ and since $\psi_1$ is a continuous in $\boldsymbol{\theta}$, therefore $\exists \, M_{r_0}$ s.t. $\|\psi_1(Y_{i,j},d_i,\boldsymbol{\theta})-\psi_1(Y_{i,j},d_i,\boldsymbol{\theta_0})\|\leq M_{r_0}$. Hence, each $\psi_1(Y_{i,j},d_i,\boldsymbol{\theta})$ is also bounded in that compact neighbourhood. So, $\phi$ is bounded. Now,
  $$\frac{\partial \phi}{\partial \boldsymbol{\theta}}=
-\frac{\partial \mu(d_i,\boldsymbol{\theta})}{\partial \boldsymbol{\theta}}\psi_2(d_i,\boldsymbol{\theta}) +\psi_1(Y_{i,j},d_i,\theta)\frac{\partial \psi_2(d_i,\boldsymbol{\theta})}{\partial \boldsymbol{\theta}}$$
  The weights functions are so chosen that $\psi_2$ is continuously differentiable. Then $\psi_2$ and its' derivative is bounded in the compact interval. Using similar reasoning as above, one can also show the derivative of $\phi$ is bounded. Thus $H3$ is satisfied for the above choice of $\phi$.
  
  \item[H4:] Under the true dose-response model, $\widehat{\boldsymbol{\theta}}_{n,0}$ are estimates of the un-wegihted non-linear regression, so it follows, $\widehat{\boldsymbol{\theta}}_{n,0} \to \boldsymbol{\theta_0}$ a.s.
  \item[N1:] Follows from the arguments for H3.
 \end{itemize}
 Thus Theorem \ref{Theorem1} and Theorem \ref{Theorem2} are satisfied.
 \newpage
\subsection{Weighted Regression using IRWLS}
\label{Appendix A.3}
The following algorithm elaborates how the IRWLS is executed in our approach and how the optimal estimates are obtained.

\begin{algorithm}
\caption{Algorithm for estimating MED by IRWLS}
\label{euclid}
\begin{algorithmic}[1]

\State $iteration \gets 1$
\State $converged \gets FALSE$
\State $Set \, tolerance \, for \, convergence \, in \, tol$
\BState \emph{loop}:
 \While{$iteration < maxit$}
  \If{$iteration =1$} 
   \State $weights= 1$ $\forall$ dose 
  \Else
   \State $weights=w_i, \, \text{for some weight function in Section \ref{Weights}}$ 
   \If{model being fitted belongs to linear family}
     \If{model=linear} X=[1;dose]
     \ElsIf{model=linlog} X=[1;log(dose)] 
     \EndIf
    \State Solve for  $WY=WX\beta$ using QR decomposition and obtain $\hat{\beta}$.
  \Else
   \State $bnds\gets default\, band\, limits$ (Dose Finding Package)
   \State Grid search (for $\gamma$ in equation (\ref{anovamodel}) of Section \ref{Parametric set-up}) in the above 
   \State band limit to obtain the $\hat{\gamma}$ which minimizes the SSE. Refine the \State bounds of $\gamma$ around the estimate obtained by grid search
   \Procedure{How to refine bounds?}{}
     \State $strt \gets estimate\, obtained\, from\, grid\, search$
     \State $bnds\gets default\, band\, limits$
     \State $N \gets 30$
     \State $dif \gets \frac{bnds[2]-bnds[1]}{N}$
     \State $bnds[1] \gets \max(c(strt-1.1*dif), bnds[1])$
     \State $bnds[2] \gets \min(c(strt+1.1*dif), bnds[2])$
   \EndProcedure
    \State With the refined bounds get the optimal estimates of the parameters
    \State  of the full model, renew weights and check for convergence
  \EndIf

  \If{$\mathrel{\big(}\frac{(\widehat{MED}_{new}-\widehat{MED}_{old})}{\widehat{MED}_{old}}\mathrel{\big)}^2 <=tol$
  or if $\widehat{MED} <=0$}
  \State $converged \gets TRUE$  \Comment{Condition for convergence}
  \State
  \Return $\hat{\beta}$ and $\widehat{MED}$  
  \Else
  \State $iteration \gets iteration+1$, \textbf{goto} \emph{loop}
  \EndIf
  \EndIf
\EndWhile
\If {$iteration = maxit$ AND $converged = FALSE$}  
\State \Return{$\hat{\beta}$= estimates obtained in the first iteration}
\EndIf
\end{algorithmic}
    \end{algorithm}
\newpage
\subsection{Boxplot distribution of MED}
\label{Appendix A.4}
\begin{figure}[h]
 \centering
\subfigure[MED estimation for data simulated from the sigEmax model and fitted with the linear and sigEmax model for sample size 25]{%
\includegraphics[scale=0.55]{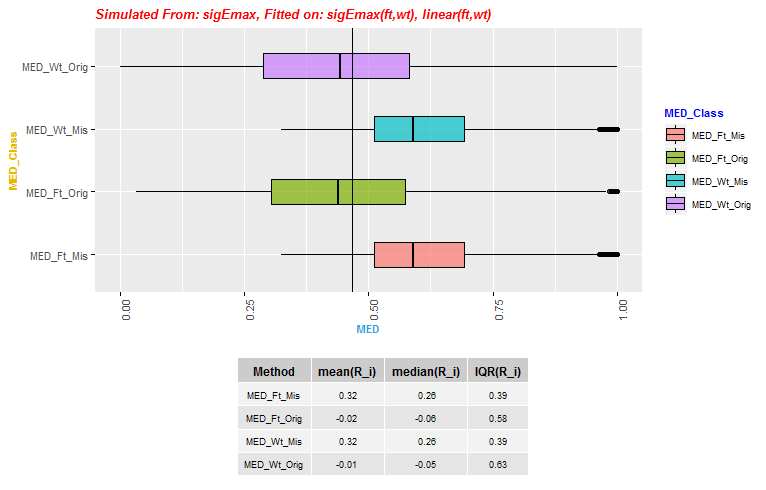}}
\qquad
\subfigure[MED estimation for data simulated from the sigEmax model and fitted with the linear and sigEmax model for sample size 50]{%
\includegraphics[scale=0.55]{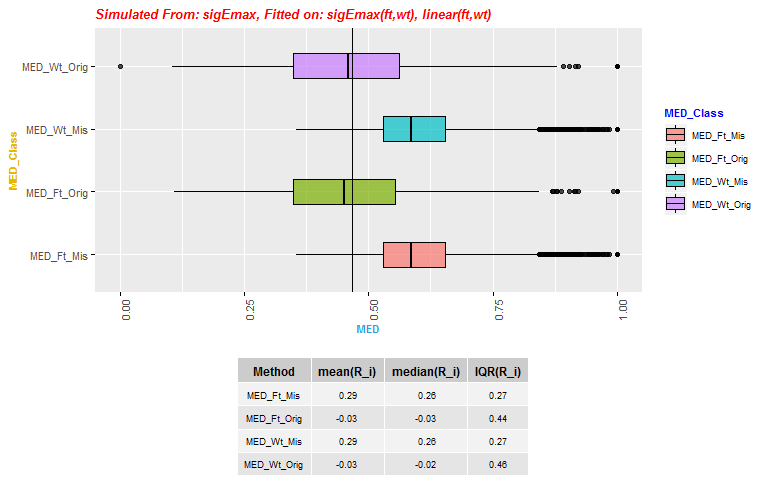}}
\caption{Boxplot distribution of MED estimated under the classical and RR approach for weight $w_4$. This is followed by a table which summarizes the mean, median and IQR of $R_i$ for the 5000 simulation trials. Figure (a) and Figure (b) show the MED estimated with the true model (sigEmax) and the misspecified model (linear) for the data simulated from the sigEmax model in Table \ref{tab:SimulationTable} with sample size $25$ and $50$, respectively. \texttt{MED\_Ft\_Orig} and \texttt{MED\_Ft\_Mis} denote the MED estimated using the classical approach with the true and misspecified model, respectively. \texttt{MED\_Wt\_Orig} and \texttt{MED\_Wt\_Mis} denote the MED estimated using the RR approach with the true and misspecified model, respectively.}
 \label{fig:RRsigEmaxLin2}
\end{figure}
\begin{figure}[h]
 \centering
\subfigure[MED estimation for data simulated from the sigEmax model and fitted with the linear and sigEmax model for sample size 25]{%
\label{fig:sigEmaxlinear25}%
\includegraphics[scale=0.50]{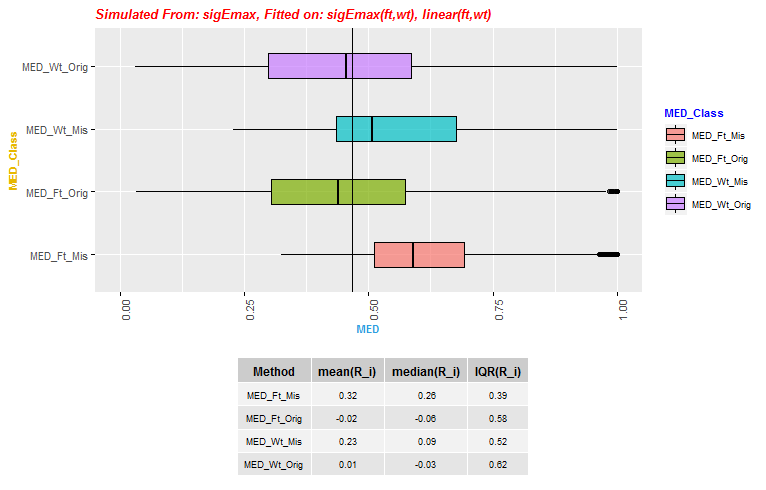}}
\qquad
\subfigure[MED estimation for data simulated from the sigEmax model and fitted with the linear and sigEmax model for sample size 50]{%
\label{fig:sigEmaxlinear50}%
\includegraphics[scale=0.50]{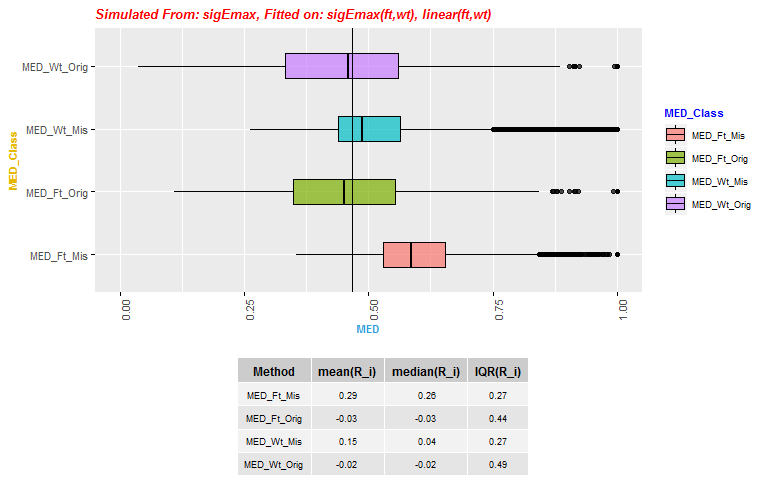}}
\caption{Boxplot distribution of MED estimated under the classical and RR approach for weight $w_5$. This is followed by a table which summarizes the mean, median and IQR of $R_i$ for the 5000 simulation trials. Figure (a) and Figure (b) show the MED estimated with the true model (sigEmax) and the misspecified model (linear) for the data simulated from the sigEmax model in Table \ref{tab:SimulationTable} with sample size $25$ and $50$, respectively. \texttt{MED\_Ft\_Orig} and \texttt{MED\_Ft\_Mis} denote the MED estimated using the classical approach with the true and misspecified model, respectively. \texttt{MED\_Wt\_Orig} and \texttt{MED\_Wt\_Mis} denote the MED estimated using the RR approach with the true and misspecified model, respectively.}
 \label{fig:RRsigEmaxLin3}
\end{figure}
\begin{figure}[h]
 \centering
\subfigure[MED estimation for data simulated from the sigEmax model and fitted with the emax and sigEmax model for sample size 25]{%
\label{fig:sigEmaxEmax31}%
\includegraphics[scale=0.55]{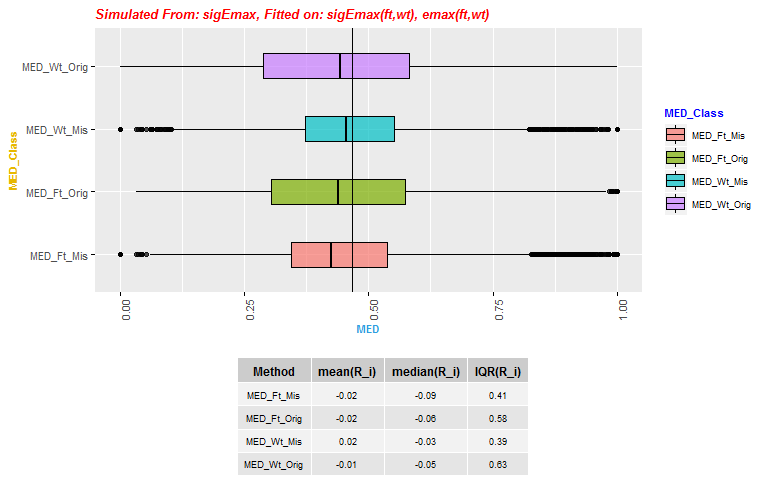}}
\qquad
\subfigure[MED estimation for data simulated from the sigEmax model and fitted with the emax and sigEmax model for sample size 50]{%
\label{fig:sigEmaxEmax32}%
\includegraphics[scale=0.55]{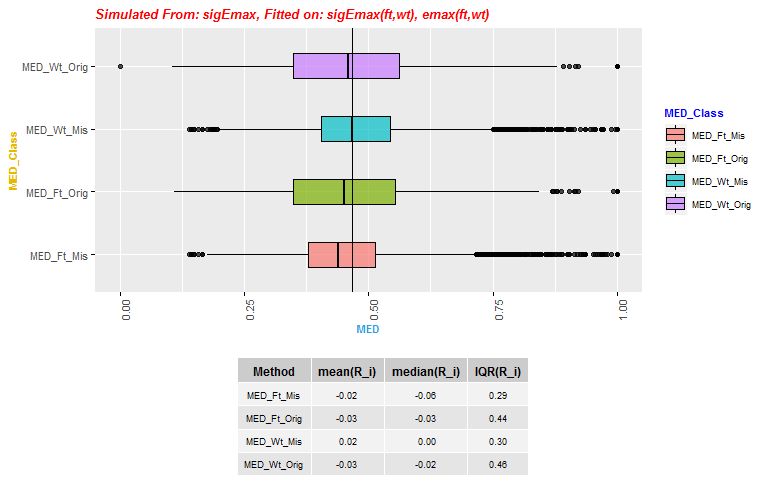}}
\caption{Boxplot distribution of MED estimated under the classical and RR approach for weight $w_4$. This is followed by a table which summarizes the mean, median and IQR of $R_i$ for the 5000 simulation trials. Figure (a) and Figure (b) show the MED estimated with the true model (sigEmax) and the misspecified model (emax) for the data simulated from the sigEmax model in Table \ref{tab:SimulationTable} with sample size $25$ and $50$, respectively. \texttt{MED\_Ft\_Orig} and \texttt{MED\_Ft\_Mis} denote the MED estimated using the classical approach with the true and misspecified model, respectively. \texttt{MED\_Wt\_Orig} and \texttt{MED\_Wt\_Mis} denote the MED estimated using the RR approach with the true and misspecified model, respectively.}
 \label{fig:RRsigEmaxEmax3}
\end{figure}
\begin{figure}[h]
 \centering
\subfigure[MED estimation for data simulated from the sigEmax model and fitted with the emax and sigEmax model for sample size 25]{%
\label{fig:sigEmaxEmax21}%
\includegraphics[scale=0.55]{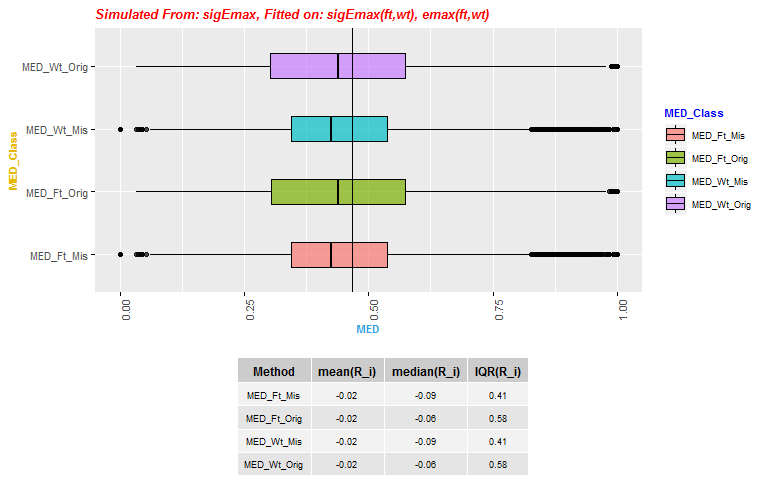}}
\qquad
\subfigure[MED estimation for data simulated from the sigEmax model and fitted with the emax and sigEmax model for sample size 50]{%
\label{fig:sigEmaxEmax22}%
\includegraphics[scale=0.55]{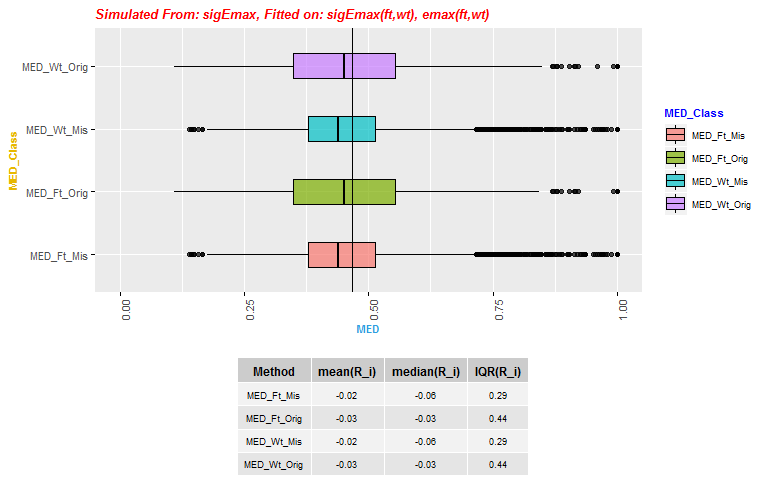}}
\caption{Boxplot distribution of MED estimated under the classical approach and weighted regression method (RR) for weight $w_6$. This is followed by a table which summarizes the mean, median and IQR of $R_i$ for the 5000 simulation trials. Figure (a) and Figure (b) show the MED estimated with the true model (sigEmax) and the misspecified model (emax) for the data simulated from the sigEmax model in Table \ref{tab:SimulationTable} with sample size $25$ and $50$, respectively. \texttt{MED\_Ft\_Orig} and \texttt{MED\_Ft\_Mis} denote the MED estimated using the classical approach with the true and misspecified model, respectively. \texttt{MED\_Wt\_Orig} and \texttt{MED\_Wt\_Mis} denote the MED estimated using the RR approach with the true and misspecified model, respectively.}
 \label{fig:RRsigEmaxEmax2}
\end{figure}

%

\let\oldsection\section
\renewcommand\section{\clearpage\oldsection}
\bibliographystyle{apalike}  
\bibliography{references}

\end{document}